\RequirePackage{fix-cm}
\documentclass[onecolumn, a4paper, 12pt]{article}

\usepackage{graphicx}
\usepackage{mathptmx}
\usepackage{latexsym}
\usepackage{graphics}
\usepackage{epsfig}
\usepackage{natbib}
\usepackage{times}
\usepackage{epstopdf}
\usepackage{subfigure}
\usepackage{array}
\usepackage{float}
\usepackage[margin=1.0in]{geometry}
\usepackage{amsmath,amsfonts,amssymb,epsfig,subfigure}
\usepackage{amsmath,amsfonts,amssymb}
\usepackage{setspace}
\usepackage{authblk}
\usepackage{multirow}
\usepackage{colortbl}
 \usepackage{ifpdf}
 \ifpdf     
    \else
      \fi

\bibpunct{[}{]}{,}{n}{,}{,}
\DeclareMathSizes{10}{9}{7}{5}
\DeclareGraphicsExtensions{.pdf}
\renewcommand{\le}{\leqslant}

\newcommand{\be}{\begin{equation}}
\newcommand{\en}{\end{equation}}

\newcommand{\tr}{\textrm{tr}}

\newcommand{\ee}{\textrm{e}}
\newcommand{\Keywords}[1]{\par\noindent{\small{\em Keywords\/}: #1}}

\begin{document}

\title{Automated Estimation of Collagen Fibre Dispersion in the Dermis and its Contribution to the Anisotropic Behaviour of Skin}

\author[1,2,3]{Aisling N\'{i} Annaidh }
\author[4]{Karine Bruy\`{e}re}
\author[5,1]{Michel Destrade}
\author[1]{Michael D. Gilchrist}
\author[2,3]{Corrado Maurini}
\author[4]{Melanie Ott\'{e}nio}
\author[7]{Giuseppe Saccomandi}


\affil[1]{School of Mechanical {\&} Materials Engineering, University College Dublin, Belfield, Dublin 4, Ireland}
\affil[2]{UPMC, Univ Paris 6, UMR 7190, Institut Jean Le Rond d'Alembert, Bo\^ite courrier 161-2, 4 Place Jussieu, F-75005, Paris France}     
\affil[3]{CNRS, UMR 7190, Institut Jean Le Rond d'Alembert, Bo\^ite courrier 161-2, 4 Place Jussieu, F-75005, Paris France}           
\affil[4]{Universit\'{e} de Lyon, F-69622, Lyon, France, Ifsttar, LBMC, UMR\_T9406, F-69675, Bron, Universit\'{e} Lyon 1, Villeurbanne}
\affil[5]{School of Mathematics, Statistics and Applied Mathematics, National University of Ireland Galway, Galway, Ireland}
\affil[6]{School of Human Kinetics, University of Ottawa, Ontario K1N 6N5, Canada}
\affil[7]{Dipartimento di Ingegneria Industriale, Universit\`a degli Studi di Perugia, 06125 Perugia, Italy}

\date{}
\maketitle
\pagebreak

\begin{abstract}
Collagen fibres play an important role in the mechanical behaviour of many soft tissues. Modelling of such tissues now often incorporates a collagen fibre distribution.  However, the availability of accurate structural data has so far lagged behind the progress of anisotropic constitutive modelling. Here, an automated process is developed to identify the orientation of collagen fibres using inexpensive and relatively simple techniques. The method uses established histological techniques and an algorithm implemented in the MATLAB image processing toolbox. It takes an average of 15 seconds to evaluate one image, compared to several hours if assessed visually. The technique was applied to histological sections of human skin with different Langer line orientations and a definite correlation between the orientation of Langer lines and the preferred orientation of collagen fibres in the dermis (P$<$0.001, R$^{2}$= 0.95) was observed.
The structural parameters of the Gasser-Ogden-Holzapfel (GOH) model were all successfully evaluated. The mean dispersion factor for the dermis was $\kappa $ = 0.1404 $\pm $ 0.0028. The constitutive parameters $\mu$, $k_{1}$ and $k_{2}$ were evaluated through physically-based, least squares curve-fitting of experimental test data. The values found for $\mu$, $k_{1}$ and $k_{2}$ were 0.2014 MPa, 243.6 and 0.1327, respectively.
Finally, the above model was implemented in ABAQUS/ Standard and a finite element (FE) computation was performed of uniaxial extension tests on human skin. 
It is expected that the results of this study will assist those wishing to model skin, and that the algorithm described will be of benefit to those who wish to evaluate the collagen dispersion of other soft tissues.


\end{abstract}
\pagebreak

\section{Introduction}
\label{intro}

Collagen fibres govern many of the mechanical properties of soft tissues, in particular their anisotropic behaviour. Due to the complex nature of fibre arrangements, such tissues are often represented as either isotropic or transversely isotropic \citep{Evans09, Bischoff00, Bischoff04}. The availability of quantitative structural data on the orientation and concentration of collagen fibres is crucial in order to describe the behaviour of these soft tissues accurately.

A number of studies have used statistical distributions to describe the fibre arrangement in soft tissues. \citet{Lanir83} was the first to attempt to account for fibre dispersion. Lanir's method expresses the mechanical response in terms of angular integrals. This technique accounts for the contribution of infinitesimal fractions of collagen fibres in a particular orientation. This approach leads to accurate results but it is not a practical option for efficient numerical implementation \citep{Cortes10}. The second approach uses generalised structure tensors (GST), which are assumed to represent the three-dimensional distribution of the fibres. 
The strain energy function is then calculated by using the average stretch rather than by using a stretch for each individual fibre. 
GST is a simpler method than Lanir's, and is easily implemented in FE algorithms \citep{Gasser06}.

In this paper we implement the GST method proposed by \citet{Gasser06}, described in detail in Section~\ref{Gasser}. 
In that model, structural parameters such as the fibre dispersion parameter and mean orientation of fibres are required. 
In order to quantify the orientations of collagen fibres in the human dermis an automated process is developed for detecting collagen orientations in histology slides.

Histology is the microscopic study of cells and tissue. It is an important diagnostic tool in medicine which is utilised here for the purpose of analysing collagen orientation. Traditionally, histology slides are examined by expert observers who assess individual slides visually. This method is both time consuming and subjective. An automated process would enable large volumes of images to be analysed quickly and improve the objectivity of the task. Much research has been conducted recently on imaging techniques of biological soft tissue for the extraction of structural data; however these techniques require either expensive equipment or manual post processing \citep{Flamini10, Verhaegen10, Wu03,Yasui04}. There has been little research to date carried out on the automated analysis of histology slides. Four notable exceptions are the work of \citet{VanZuijlen02}, \citet{Noorlander02}, \citet{Elbischger04} and \citet{Jor11}. 

\citet{Elbischger04} successfully automated the analysis of collagen fibre orientation in the human adventitia, (the outermost membrane of the artery). Their technique was based on the automated analysis of Transmitted Light Microscopy images which were stained with Elastica van Gie\-son. The algorithm uses a ridge and valley analysis to detect the orientation of collagen fibres and segments regions of homogeneous fibre orientations. This is a somewhat complicated technique which requires substantial coding to implement. There are also assumptions made which cannot be applied to the dermis, i.e. that the image contains at least 50\% collagen and that the fibres have a common orientation. In contrast, \citet{Noorlander02} used a relatively simple technique upon which our own algorithm is based. In their study, Picrosirius red staining was used together with epipolarised light to image the fibres. Individual collagen fibres were detected and ellipses fit to the longest ten fibres. While their technique is quantitative it is not an automated technique and requires the user to identify fibre orientations visually. The results refer to the `collagen alignment index' which is the mean length of the best fit ellipse and has no connection with the angular orientation of fibres, which itself is not measured. \citet{VanZuijlen02} used Fourier analysis to measure the level of anisotropy of collagen in the skin. Those authors concluded that analysis of `orientation index' by Fourier analysis is superior to conventional techniques. 
However, the `orientation index' is a measure of the anisotropy of the matrix but, just like the `collagen alignment index' of  \citet{Noorlander02}, it fails to provide information on the mean orientation of the fibres. In a recent publication \citet{Jor11} modelled the orientation of collagen fibres in porcine skin. Their technique involved the staining of porcine skin sections with Picrosirius red and image acquisition by confocal laser scanning microscopy. Their study provided quantitative results on the orientation of collagen fibres in porcine skin, however the plane of interest was normal to the epidermis and it is generally believed that the preferred orientation of collagen fibres is parallel to the epidermis \citep{Holzapfel01}. 

The chief advantage of the technique proposed here over other automated techniques is that it is an inexpensive and relatively simple technique. It can be easily implemented in MATLAB using the Image Processing Toolbox so that the algorithm can be easily amended as required for the user's specific application. It is a fully automated process (aside from the image acquisition phase) and is capable of analysing a greater number of images, and faster, than a manual analysis. It also eliminates the subjectivity which is present using traditional methods. The technique provides quantitative data on both the collagen orientation and the level of anisotropy in the dermis, but it could be easily amended to identify any tissues/objects within other soft tissues. 

Due to the anisotropic properties of collagen based soft tissues \citep{NiAnnaidh11a}, modelling now often incorporates a collagen fibre distribution. The accuracy of these structural models rely heavily on knowledge of collagen fibre orientations. While this data has recently been published for porcine skin by \citet{Jor11}, to the best of the authors' knowledge, quantitative data on the orientation of collagen fibres in the human dermis has not been previously published. With the quantitative data obtained here, structural parameters such as $\gamma $ and $\kappa $ from the Gasser-Ogden-Holzapfel model \citep{Gasser06} can be evaluated independently. This structural data, coupled with the mechanical testing of the same skin samples \citep{NiAnnaidh11a} provides us with sufficient data to model human skin using the GOH model.


\section{Materials {\&} Methods}


\subsection{Tensile tests and histology}

In vitro tensile tests of human skin were performed in Ifsttar (Institut fran{\c c}ais des sciences et technologies des transports, de l'am\'{e}nagement et des r\'{e}seaux), France. French law allows human corpses that have been donated to science to be used for research purposes. The ethics committee within Ifsttar approved the use of human biological material. 

The tensile tests were performed using a Universal Tensile Test machine at a strain rate of 0.012s$^{-1}$. The tensile load was measured with a 1kN piezoelectric load cell and the strain was measured via a displacement actuator. Skin biopsies were excised adjacent to the tensile test samples and stained with Van Gieson to dye collagen fibres pink/red. Further details of the tensile test experimental procedure and histology protocol are outlined in \citet{NiAnnaidh11a}.


\subsection{Automated detection of fibre orientations}

\begin{figure}[!ht]
\centering
\subfigure[Original histology slide. Scale bar is 1mm.]{ \includegraphics[width=0.25\textwidth]{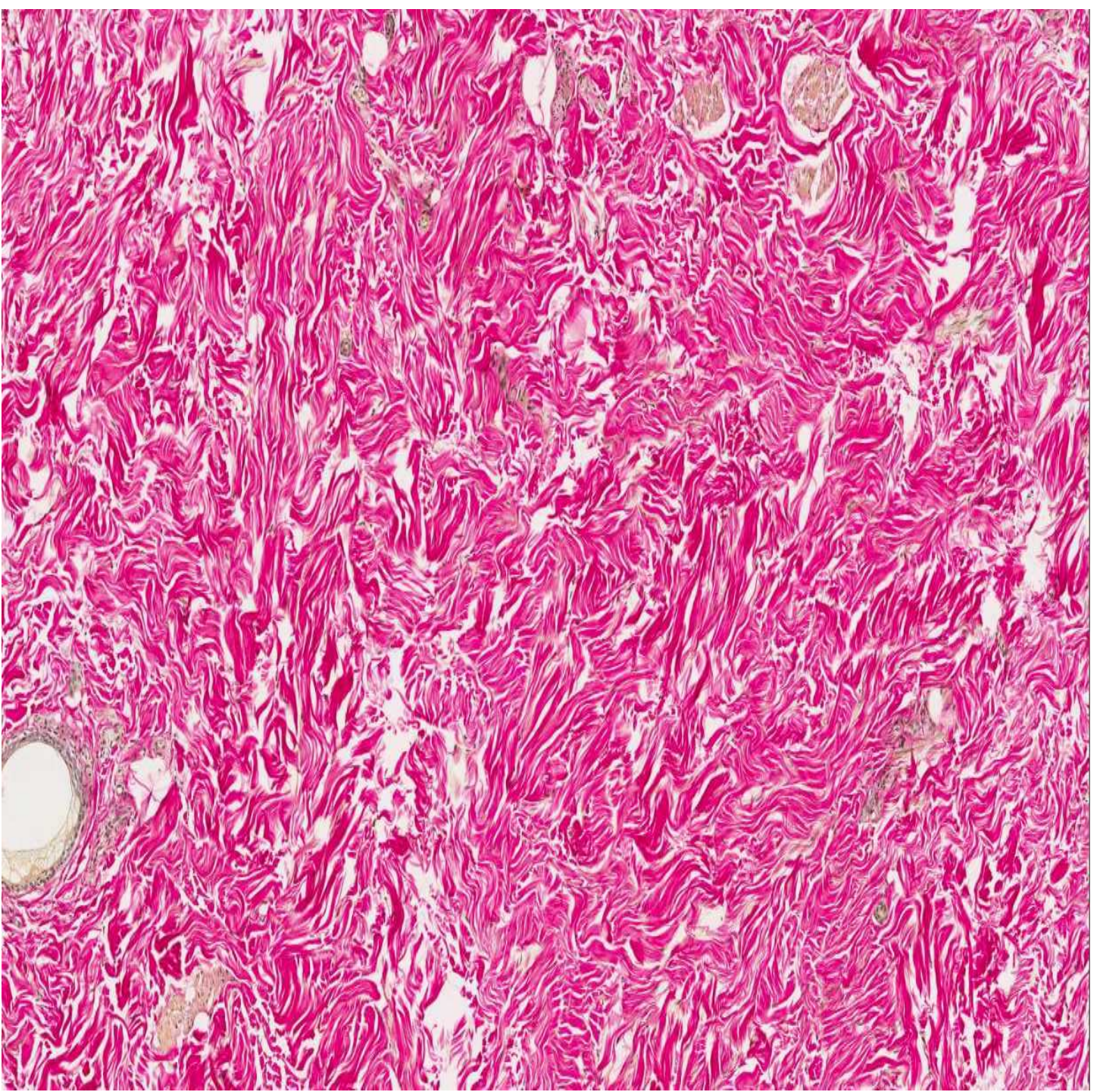}}
\subfigure[Binarised image after automated thresholding.]{ \includegraphics[width=0.25\textwidth]{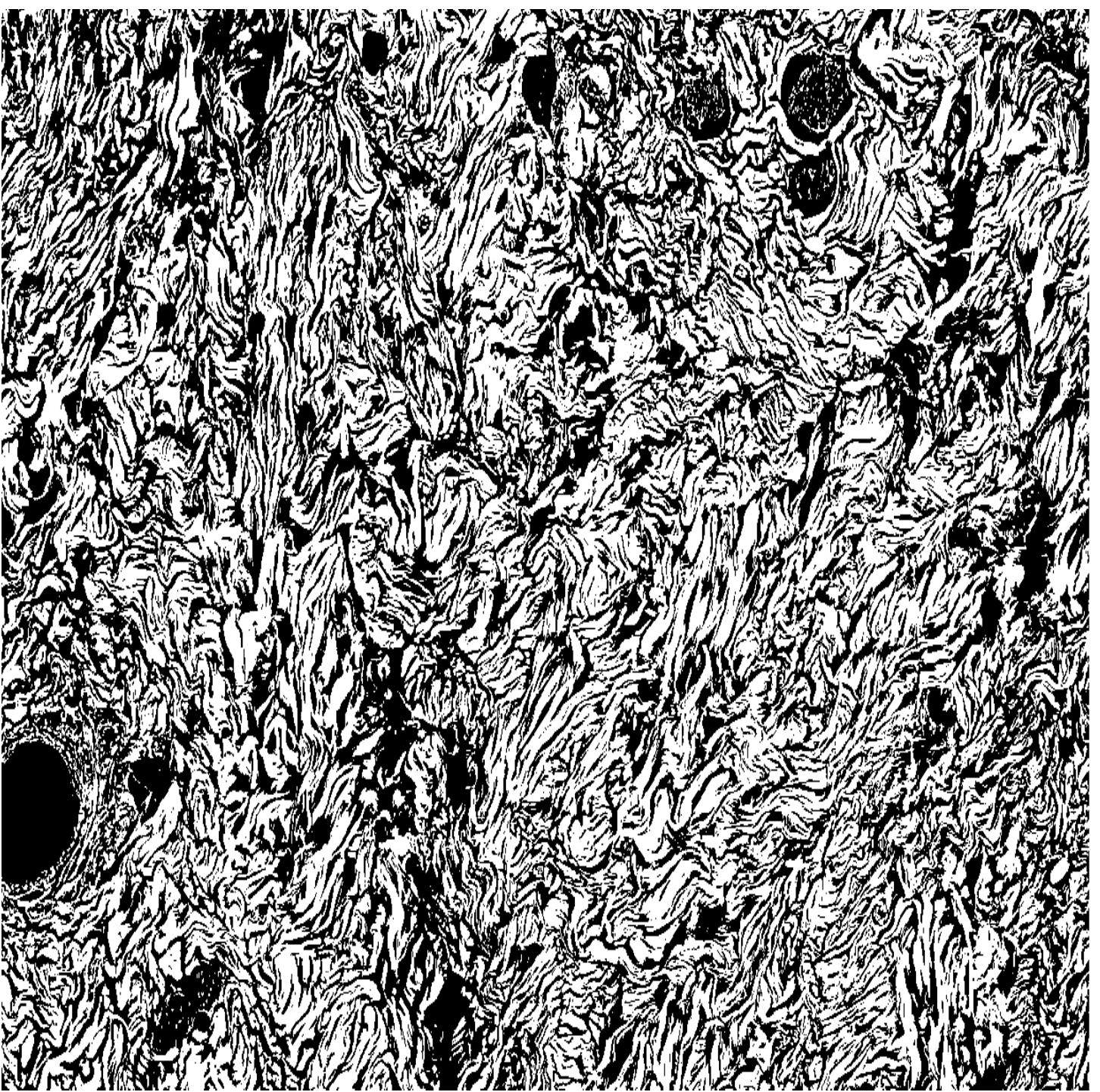}}
\subfigure[Binarised image after erosion step.]{ \includegraphics[width=0.25\textwidth]{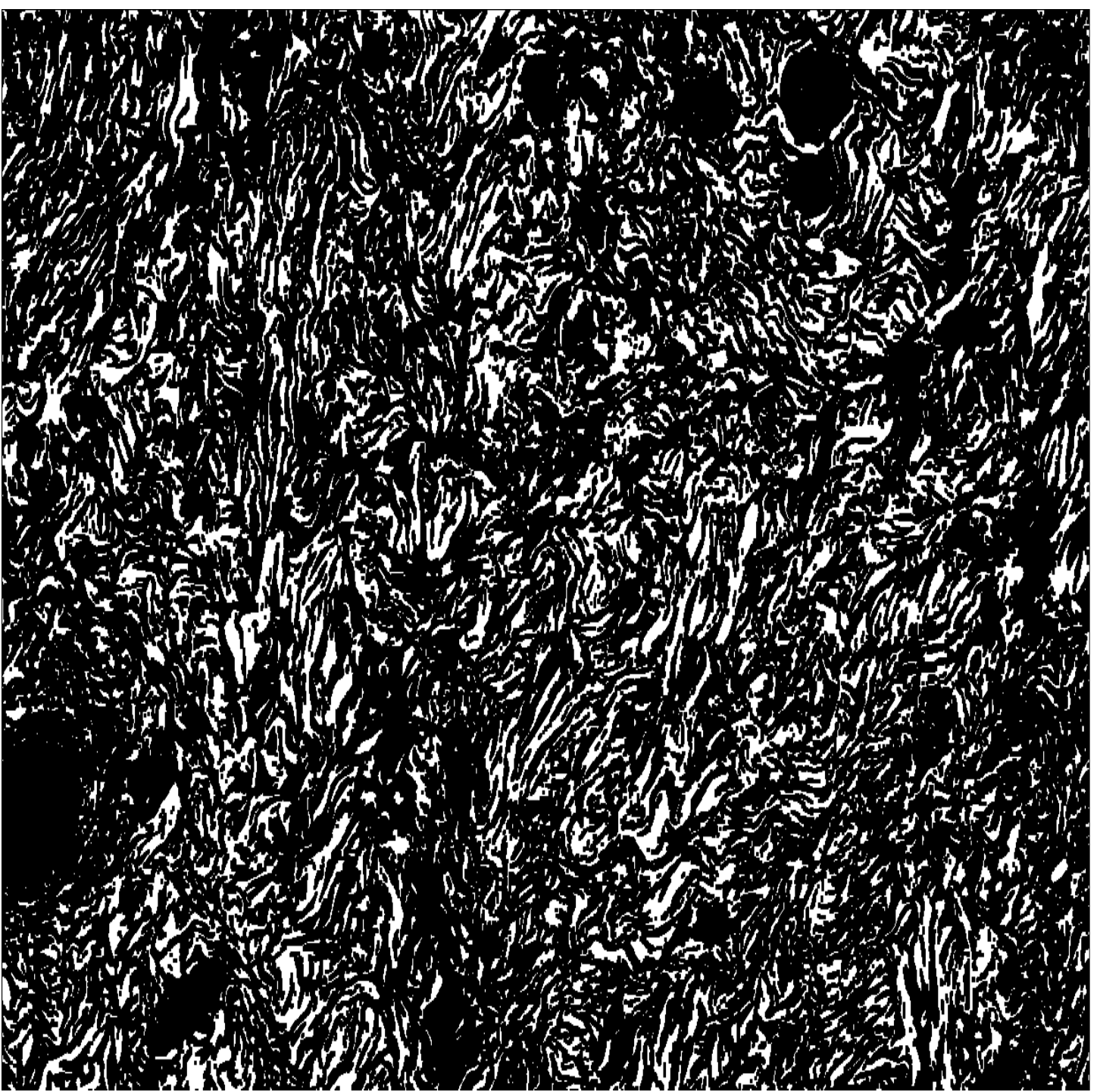}}
\subfigure[All identified collagen bundles outlined in green.]{ \includegraphics[width=0.25\textwidth]{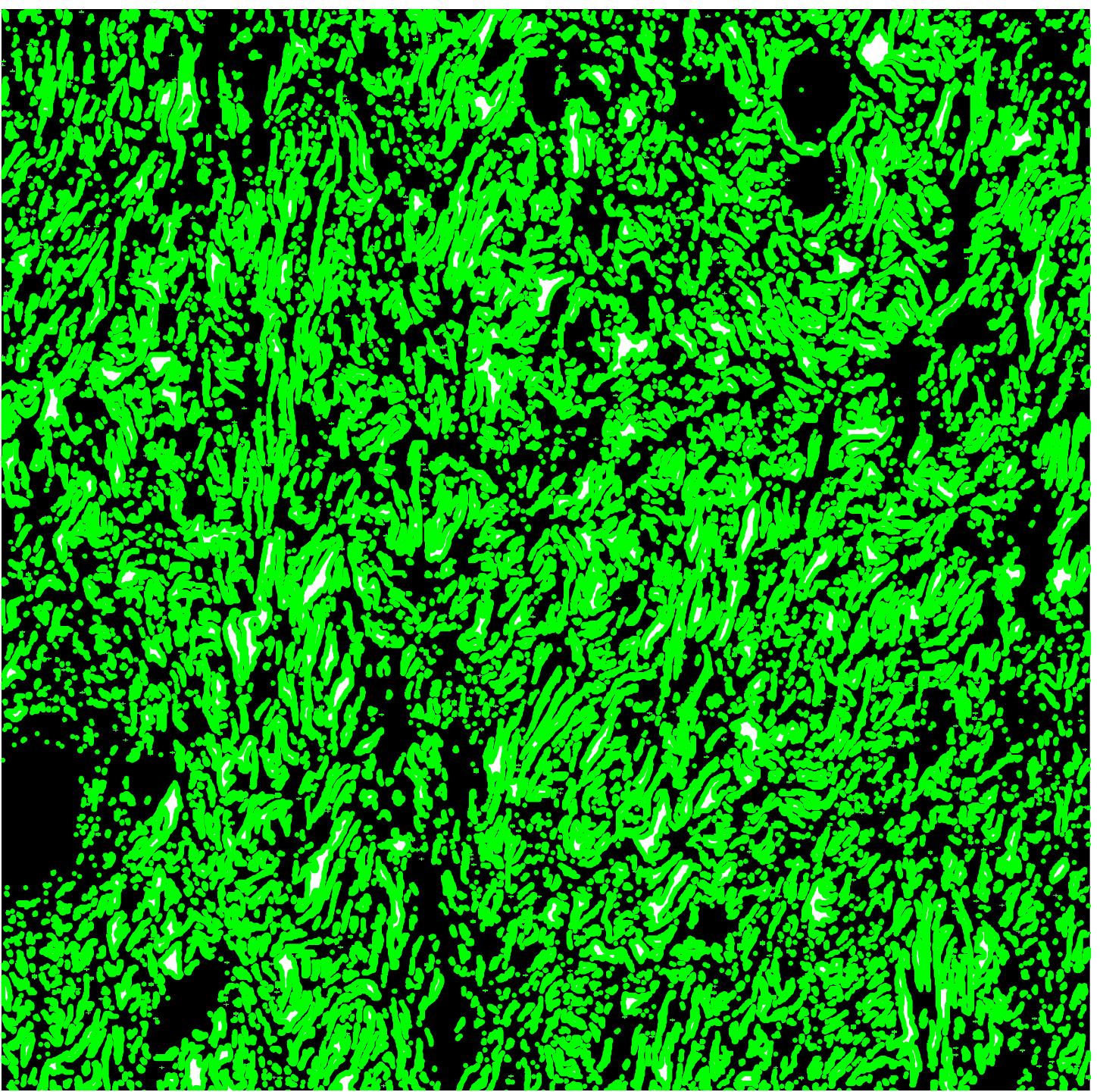}} 
\subfigure[Remaining bundles which meet area and eccentricity criteria.]{ \includegraphics[width=0.25\textwidth]{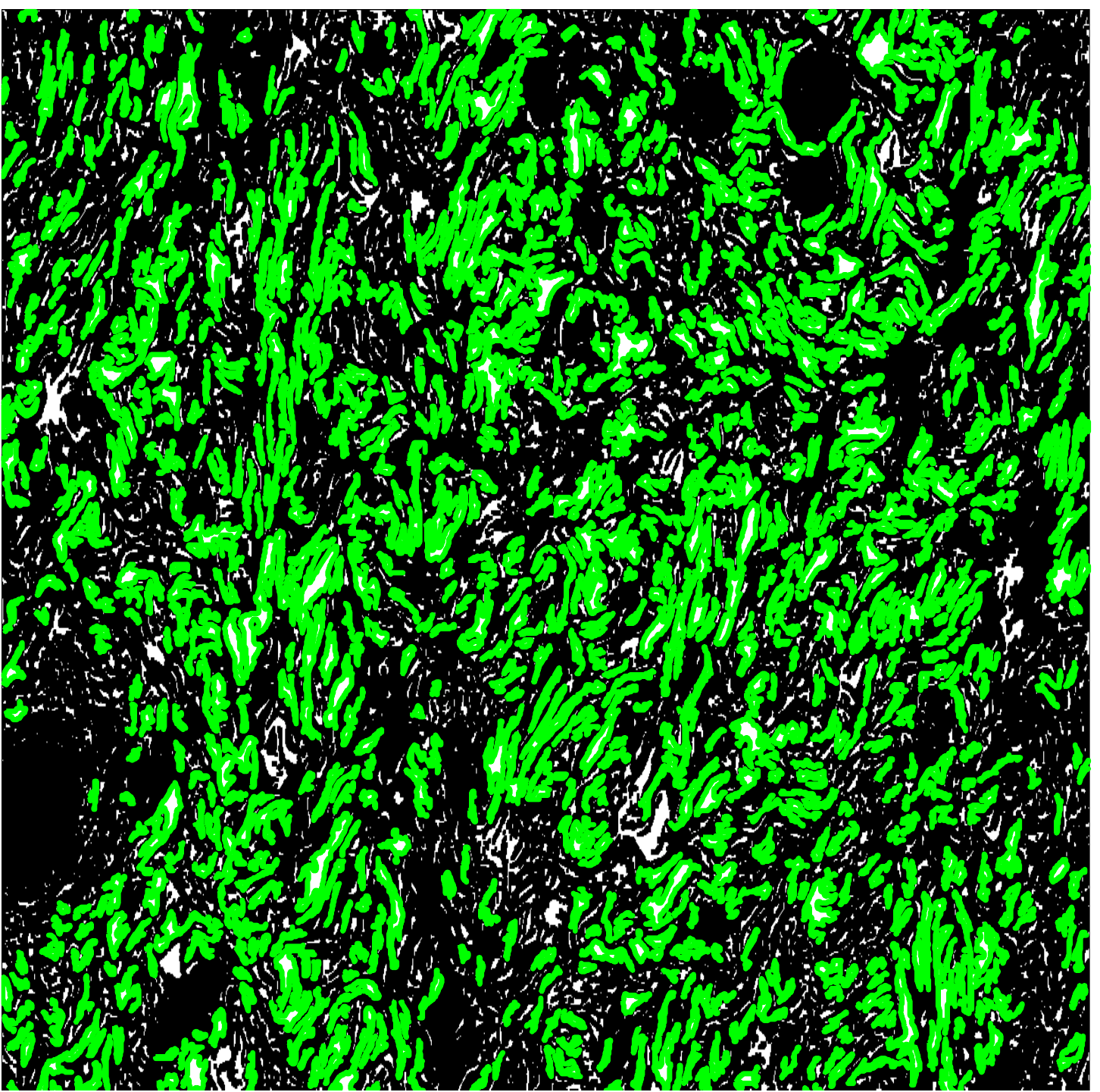}}
\subfigure[Best fit ellipse about each fibre that meets the specified criteria.]{ \includegraphics[width=0.25\textwidth]{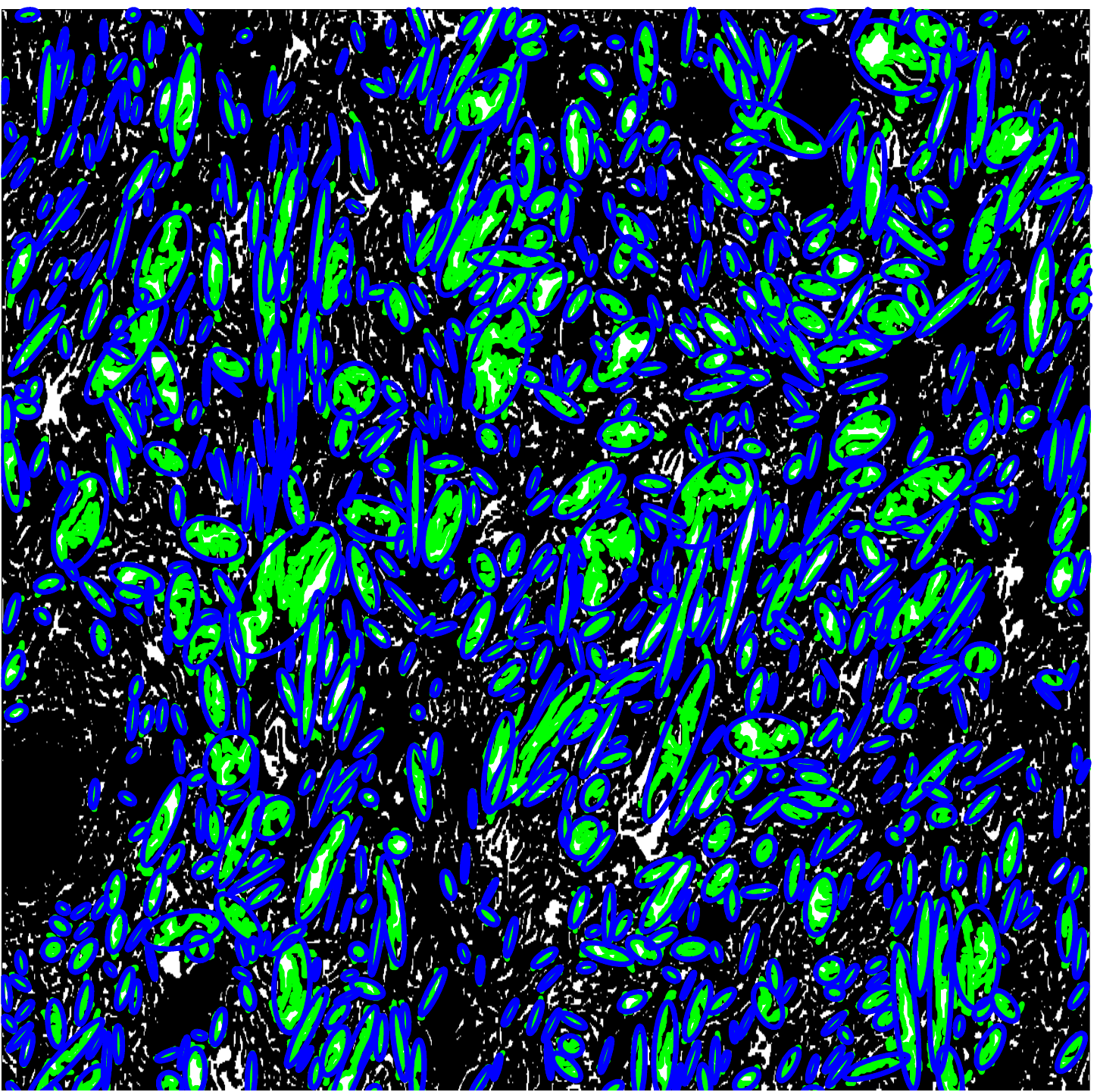}}
\caption{Images output from automated algorithm.\label{algorithm}}
\end{figure}

Images were taken of slides parallel to the epidermis (see Section~\ref{Dispersion}) using an Aperop ScanScope XT scanner and ScanScope software. This slide scanner has the ability to scan up to 120 slides automatically. The images to be analysed were selected from the reticular layer of the dermis which forms the main structural body of the skin. The optical magnification used was 5x, which is quite low; however, this was the optimal magnification to obtain a large enough field of view to be representative while also maintaining enough detail to recognise the boundaries of fibre bundles. Multiple images were taken of each sample in order to capture the entire area. The field of view for each image taken at this magnification was 2.24 mm$^{2}$. The orientation of collagen fibres were then calculated in a fully-automated customised MATLAB routine using the Image Processing Toolbox. The algorithm is described below and is illustrated in Fig.~\ref{algorithm}.

\begin{figure}[!ht]
\centering
 \includegraphics[width=0.8\textwidth]{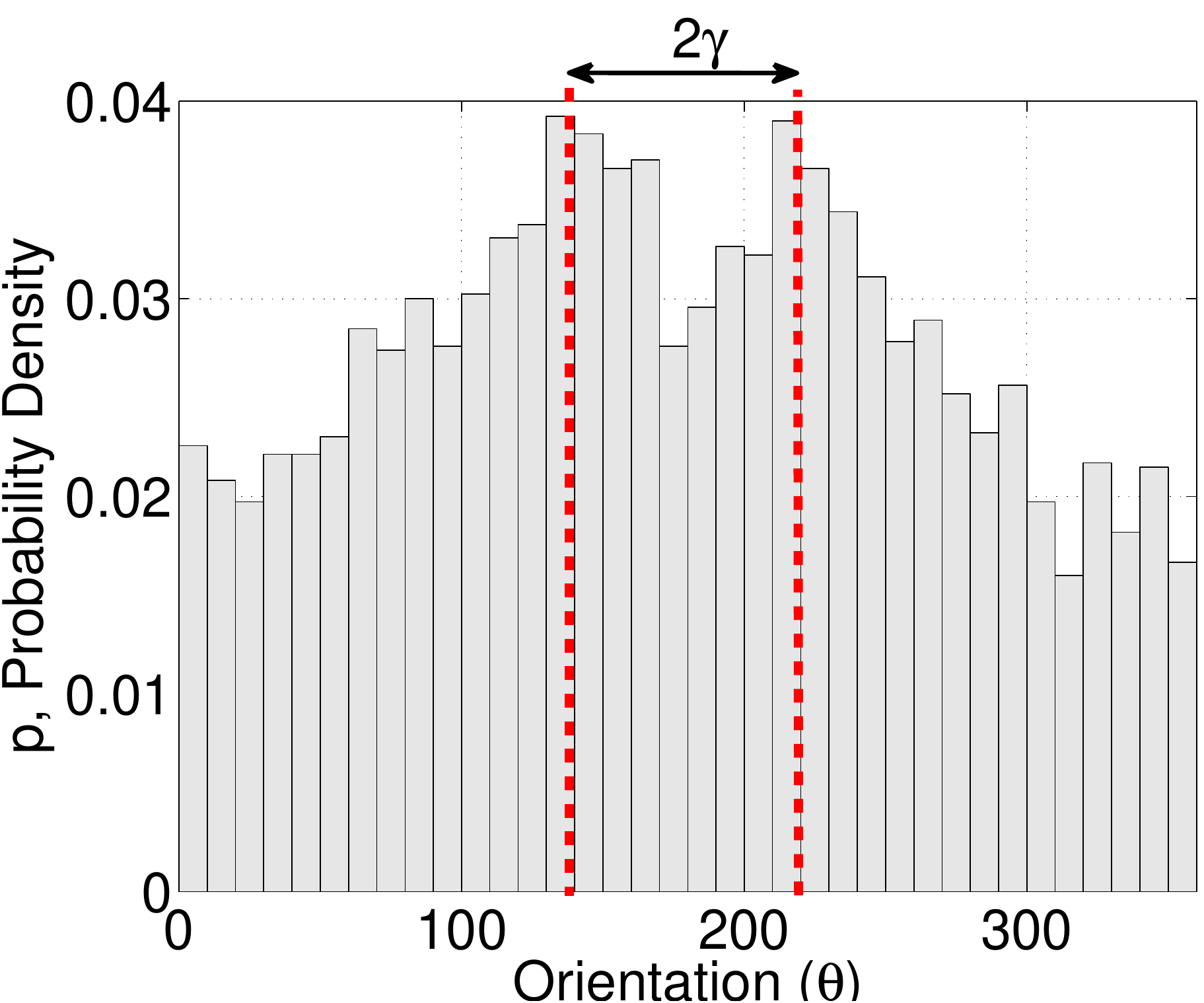}
\caption{Histogram of collagen orientations. The two distinct peaks correspond to the preferred orientation of the two fiber families. The angle, $\gamma$, is half the distance between the two peaks i.e. $\gamma$=41$^\circ$.}
\label{bimodal}
\end{figure}

\begin{itemize}
\item 
A global threshold level was computed using the \emph{graythresh} function. This function chooses a global image threshold using Otsu's method which aims to minimise the intraclass variance of black and white pixels \citep{Otsu79}. This automatically distinguishes collagen fibres from other areas which are not of interest in the analysis, such as cells or histological ground substance. The image was then binarised based on this threshold level producing an image containing black pixels for collagen and white pixels for all other areas;
\item 
Morphological operations were performed on the binary image using the \emph{bwmorph} function. An erosion step was performed to detach cross-linking fibres from each other. This step removed pixels from the boundaries using a structural element of size 3 pixels x 3 pixels. One iteration only of this step was performed which was the optimal number of iterations to detach cross-linking fibres while also leaving smaller fibres intact. 
The second morphological operation was performed using the  \emph{imfill} function. This function fills in the `holes' within the binary image, where a hole is defined as a set of isolated pixels which cannot be reached by filling in the background pixels from the edge. This step resulted in a `cleaner' image to analyse.
\item 
Individual fibre bundles are identified using \emph{bwlabel}. 
This function identifies connected components (8-connected) in a 2D binary image and labels each component individually. It uses the general procedure outlined by \citet{Haralick92};
\item 
The \emph{regionprops} function outputs a set of properties for each labelled component. The function also fits an ellipse to each component by matching the second order moments of that component to an equivalent ellipse following the procedure by \citet{Haralick92};
\item 
Only ellipses which were elongated and of a certain area were selected \footnote{Out-of-plane fibres manifest themselves as circular areas. This condition was introduced to exclude out-of-plane fibres from the analysis. A sensitivity analysis was carried out to determine the sensitivity of the algorithm to varying the area and eccentricity criteria. It was found that a $\pm $10{\%} change in both the area and eccentricity criteria led to a $\pm $1{\%} change in the mean orientation. This indicates that the selected criteria do not significantly affect the result of mean orientation. For this study it was specified that in-plane fibres must have an eccentricity larger than 0.7 and an area greater than 1000 pixels, which for our images, captured at 5x, corresponds to 5$\mu $m$^{2}$.};
\item 
The orientation of the major axis of each ellipse was calculated and taken as the approximate orientation of each component. The advantage of fitting an ellipse about each component is that the orientation is measured in a systematic and repeatable manner which can account for the non-uniformities of the shape of each component;
\item 
The orientations of each component was then plotted on a histogram (see Fig.~\ref{bimodal}). Two distinct peaks were evident from this figure. It is assumed that these two peaks correspond to the preferred orientation of two crossing families of fibres as shown in Fig.~\ref{weave}. A von Mises probability density function was fit to the data and the mean orientation and dispersion factor were calculated as described in Section~\ref{Dispersion}.
\end{itemize}


\subsubsection*{Validation}

\begin{figure}[!ht]
\centering
\includegraphics[width=0.8\textwidth]{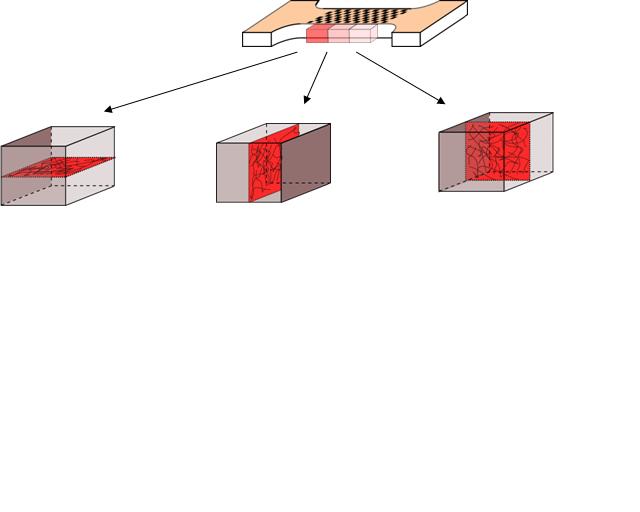}
\caption{Biopsies of skin samples for purpose of histological staining. Note that the biopsies have been sliced in three orthogonal planes.}
\label{biopsies}
\end{figure}

\begin{figure}[!ht]
\centering
 \includegraphics[width=0.5\textwidth]{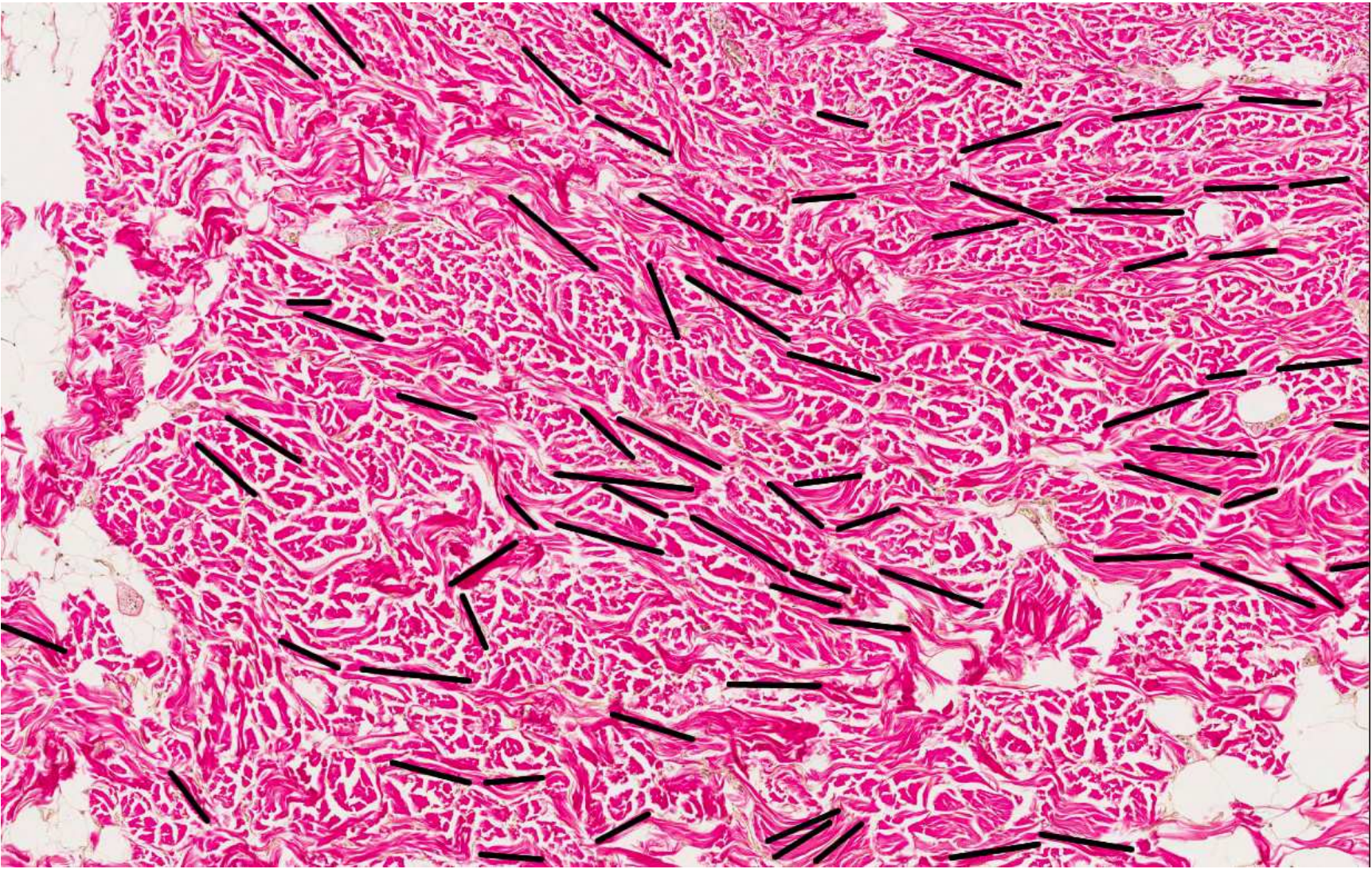}
\caption{Manual segmentation of collagen. Elongated fibres were marked by black lines, and their orientation was later measured manually.\label{manual}}
\end{figure}

As explained by \citet{VanZuijlen02}, there are difficulties involved with validating automated techniques because the \textit{true} result of each histology slide is unknown. For the validation of their algorithm, computer generated images representing collagen fibres with known orientations were used. 
However, in order to create an accurate computer generated image, a model which represents the structure of the collagen matrix is required, but no such model exists \citep{Elbischger04}. 
To validate our technique, a selection of slides in the plane perpendicular to the epidermis (see Fig.~\ref{biopsies}) were manually segmented and their mean orientation compared to those calculated through the automated process. Although it is the plane parallel to the epidermis that is used for the collection of structural data, the perpendicular plane was chosen for validation purposes because in general it displays a more orientated pattern and is easier to segment manually. Manual segmentation was performed by marking bundles of elongated fibres with a line as shown in Fig.~\ref{manual}. The orientation of each line was then measured manually and the mean calculated. 

The dermis possesses a complex interwoven collagen structure, which is difficult to assess. It is quite possible that the collagen pattern may change over a small volume. For this reason six images were taken at each depth, whereby each image spanned 3.75$mm^2$ of the 50$mm^2$ sample.
The analysis was performed at six levels of increasing depth and the mean orientation was taken as the average over these six levels. 


\subsection{Gasser-Ogden-Holzapfel Model}
\label{Gasser}

\begin{figure}[!ht]
\centering
 \includegraphics[width=1.0\textwidth]{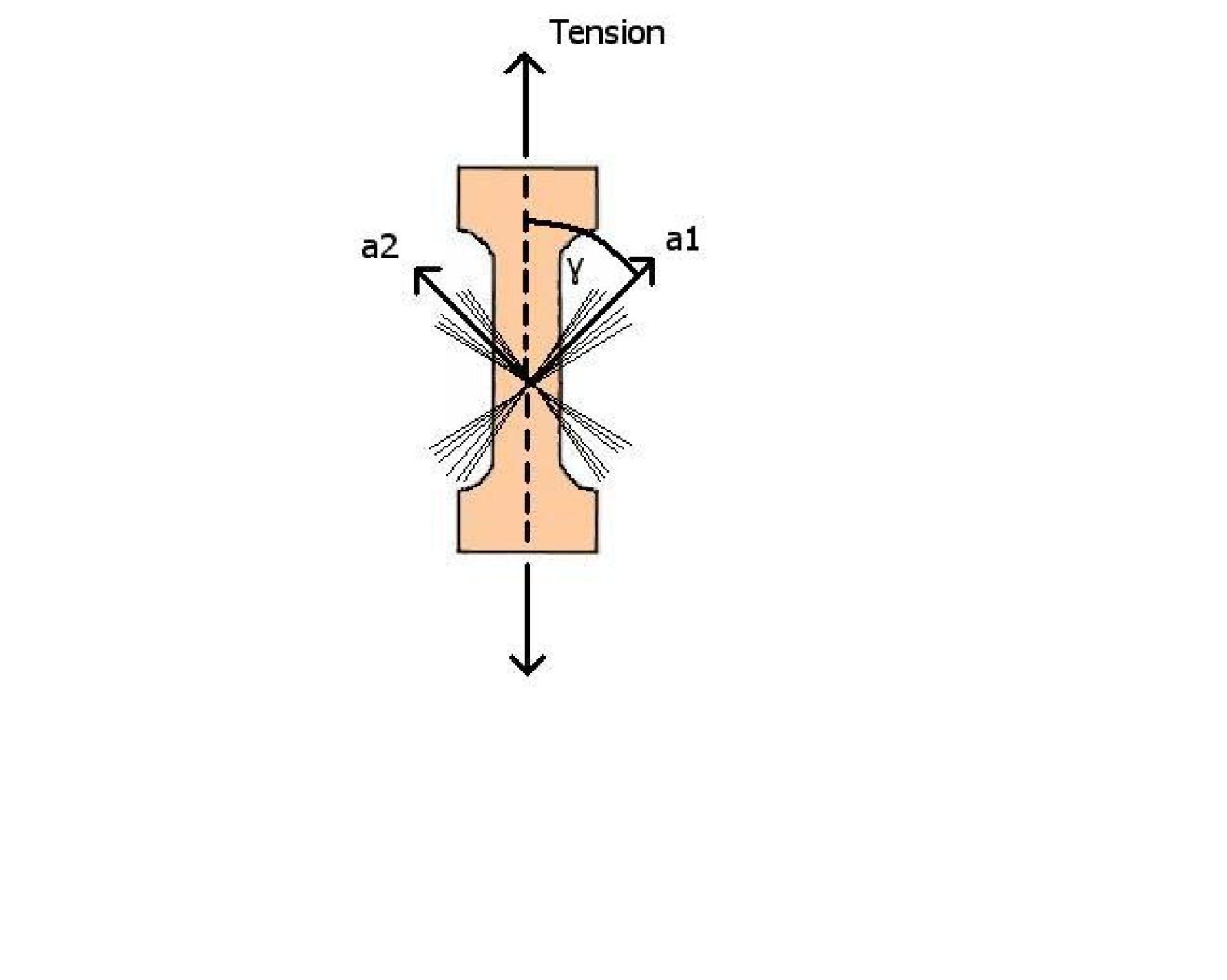}\caption{Lattice structure of crossing collagen fibres with fibre dispersion taken into account.\label{weave}}
\end{figure}

The Gasser-Ogden-Holzapfel (GOH) model applies to incompressible solids with two preferred directions aligned along the unit vectors $\mathbf{a}_{1}$ and $\mathbf{a}_{2}$ (say) in the reference configuration (see  Fig.~\ref{weave}).
Its strain energy density $\Psi$ is of the form
\begin{equation}
\Psi=\Psi(\mathbf{C},\mathbf{H}_{1},\mathbf{H}_{2})
\label{strain energy}
\end{equation}
where $\mathbf C$ is the right Cauchy-Green strain tensor, and the structure tensors $\mathbf{H}_{1}$, $\mathbf{H}_{2}$ depend on $\mathbf a_{1}$ and $\mathbf a_{2}$ and on the dispersion factors $\kappa_{1}, \kappa_{2}$ (to be detailed later), respectively, as follows
\begin{equation}
\mathbf{H}_{i}=\kappa_{i}\mathbf{I} + (1-3\kappa_{i})\mathbf{a}_{1}\mathbf{\otimes a}_2, \qquad
(i=1,2).
\label{structure tensors}
\end{equation}

Specifically, the GOH model assumes that $\Psi$ depends on $I_1 = \tr (\mathbf C)$,  $\tr(\mathbf{H}_{1}\mathbf{C})$ and $\tr(\mathbf{H}_{2}\mathbf{C})$ only, as follows
\begin{equation}
\Psi=\frac{\mu}{2}(I_{1}-3) + \mu \sum\limits_{i=1,2}\dfrac{k_{i1}}{2k_{i2}}\left\{\ee^{k_{i2}[\tr(\mathbf{H}_{i}\mathbf{C})-1]^2}-1\right\},
\label{psi}
\end{equation}
where $\mu$, $k_{i1}$, $k_{i2}$ are positive material constants, and from Eq.~\ref{structure tensors},
\begin{equation}
\tr(\mathbf{H}_{i}\mathbf{C}) = \kappa_{i}I_{i}+(1-3\kappa_{i})I_{4i},
\label{trace HC}
\end{equation}
with $I_{4i} \equiv \mathbf{a}_{i}\mathbf{\cdot C}\mathbf{a}_{i}$, two anisotropic invariants.
Note that the constitutive parameter $\mu$ has the dimensions of stress: it would be the shear modulus of the solid if there were no fibres ($k_{i1} = 0$); whilst the parameters $k_{i1}$ and $k_{i2}$ are dimensionless stiffness parameters: the $k_{i1}$ are related to the relative stiffness of the fibres in the small strain regime, and the $k_{i2}$ are related to the large strain stiffening behaviour of the fibers.

Now we focus on homogeneous uniaxial tensile tests. These can be achieved for anisotropic tissues when two families of fibres are mechanically equivalent $k_{11} = k_{21} \equiv k_{1}$ (say) and $k_{12}=k_{22} \equiv k_{2}$ (say), with the same dispersion factor $\kappa_{1}=\kappa_{2} \equiv \kappa$ (say), and when the tension occurs along the bisector of $\mathbf a_{1}$ and $\mathbf a_{2}$ (see  Fig.~\ref{weave}). 
Let us call $\gamma$ the angle between $\mathbf a_{1}$ and the tensile direction, so that now
\begin{equation}
\mathbf{a}_{1}=\cos{\gamma}\, \mathbf{i} + \sin{\gamma} \,\mathbf{j}, \qquad
\mathbf{a}_{2}=\cos{\gamma}\, \mathbf{i}-\sin{\gamma}\, \mathbf{j}.
\label{a1}
\end{equation}
Here, $\mathbf{i}$ is the unit vector in the direction of tension, and $\mathbf{j}$ is the unit vector in the lateral direction, in the plane of the sample. 
The stretch ratios along those unit vectors are $\lambda_1$ and $\lambda_2$, respectively.
Then $I_{41}=I_{42}=\lambda^{2}_{1}\cos^{2}\gamma+\lambda^{2}_{2}\sin^{2}\gamma \equiv I_{4}$ (say) and $\Psi$ reduces to
\begin{equation}
\Psi = \frac{\mu}{2} (I_{1}-3) + \mu \dfrac{k_{1}}{k_{2}}
\left\{\ee^{{k_{2}}[\kappa I_{1}+(1-3\kappa )I_{4}-1]^2} -1\right\},
\label{psireduced}
\end{equation}
giving the following expression for $\mathbf \sigma$, the Cauchy stress tensor 
\begin{equation}
\mathbf{\sigma} = 
  -p\mathbf{I}
  + 2\dfrac{\partial \Psi}{\partial I_1} \mathbf{FF}^T 
    + \dfrac{\partial \Psi}{\partial I_4}
         \left[\mathbf{Fa}_{1}\mathbf{\otimes Fa}_{1}
           + \mathbf{Fa}_{2}\mathbf{\otimes Fa}_{2}\right],
\label{cauchy}
\end{equation}
where $p$ is a Lagrange multiplier introduced by the internal constraint of incompressibility and $\mathbf{F}$ is the deformation gradient. Note that 
$\mathbf{F}\mathbf{a}_{1} \mathbf{\otimes Fa}_{1} + \mathbf{Fa}_{2} \mathbf{\otimes Fa}_{2} = 2(\lambda_{1}\cos\gamma)^2\mathbf{i \otimes i} + 2(\lambda_{2}\sin\gamma)^2\mathbf{j \otimes j}$,
showing that $\mbox{\boldmath{$\sigma$}}$ is diagonal in the $\{\mathbf{i}$, $\mathbf{j}$, $\mathbf{k}\}$ basis. 
Its components are
\begin{align}
& \sigma_{11}=-p+2(\Psi_{1}+\Psi_{4}\cos^2\gamma)\lambda^2_{1}\neq0, \notag \\
&\sigma_{22}=-p+2(\Psi_{1}+\Psi_{4}\sin^2\gamma)\lambda^2_{2}=0, \notag \\
&\sigma_{33}=-p+2\Psi_{1}\lambda^{-2}_{1}\lambda^{-2}_2=0,
\label{nonzero_sigma}
\end{align} 
where 
\begin{align}
& 2\Psi_1= \mu(1+ 4k_1\kappa\alpha \ee^{k_2\alpha^2}),\notag \\ 
& 2\Psi_4 = 4\mu k_1(1-3\kappa)\alpha \ee^{k_2\alpha^2}, \notag  \\
& \alpha = \kappa(\lambda_1^2+\lambda_2^2+\lambda_1^{-2}\lambda_2^{-2}) + (1-3\kappa)(\lambda_1^2 \cos^2\gamma + \lambda_2^2\sin^2\gamma) - 1.
\end{align}

Now, eliminate $p$ from the stress components to get the two equations
\begin{align}
& \sigma_{11} = \mu(\lambda_1^2 - \lambda_1^{-2} \lambda_2^{-2}) + 4 \mu k_1 \alpha \ee^{k_2\alpha^2} \left[\kappa(\lambda_1^2 - \lambda_1^{-2}\lambda_2^{-2}) + (1 - 3\kappa) \lambda_1^2 \cos^2\gamma \right], 
\label{sigma1} \\
& 0 = \lambda_2^2-\lambda_1^{-2}\lambda_2^{-2} + 4k_1\alpha \ee^{k_2\alpha^2} \left[\kappa (\lambda_2^2 - \lambda_1^{-2} \lambda_2^{-2}) + (1 - 3\kappa) \lambda_2^2 \sin^2\gamma \right].
\label{sigma2}
\end{align} 
 
Equation \eqref{sigma2} gives the relationship between the tensile stretch and the lateral stretch and allows, implicitly, $\lambda_2$ to be expressed in terms of $\lambda_1$. 
Substituting then into \eqref{sigma1} gives the $\sigma_{11}$--$\lambda_1$ stress-stretch relationship.
In the isotropic limit, $\kappa=1/3$ (see Section \ref{Dispersion}), and \eqref{sigma2} yields the well-known relationship $\lambda_2 = \lambda_1^{-1/2}$ for uniaxial tension in incompressible solids.

The two equations \eqref{sigma1}-\eqref{sigma2} form the basis of a numerical determination of the constitutive parameters $\mu$, $k_1$ and $k_2$, assuming that the structural parameters $\kappa$ and $\gamma$ are known. It should be noted here that the inclusion of $\mu$ in the anisotropic term of equation \eqref{psireduced} is not standard for the GOH model and has been added here for ease of calculation.
We now quantify further those latter parameters, $\kappa$ and $\gamma$ .


\subsection{Fibre Dispersion}
\label{Dispersion}

The GOH model assumes that the mean orientation of collagen fibres has no out-of-plane component. Our histological examination of the skin indicates that the majority of collagen fibres in the dermis run parallel to the epidermis. Slides parallel to the epidermis had, on average, three times less cross-sectioned fibres than slides perpendicular to the epidermis. This is in agreement with \citet{Holzapfel01} who state that the preferred orientation of the 3D collagen fiber network lies parallel to the surface, but to prevent out-of-plane shearing, some fiber orientations have components which are out-of-plane. Despite the assumption that the fibres have no out-of-plane component in the GOH model, the three-dimensional nature of the adopted distribution implies that although the preferred orientation of the fibers are in the plane parallel to the epidermis, some fibers orientations have an out-of-plane component \citep{Gasser06}.  

Here we assumed that each of the two families of collagen fibres is distributed according to a  $\pi $-periodic Von Mises distribution, which is commonly assumed for directional data. 
The standard $\pi $-periodic Von Mises Distribution is normalized and the resulting density function, $\rho(\Theta)$, reads as follows,
\begin{equation}
\rho(\Theta)=4\sqrt{\dfrac{b}{2\pi}}\dfrac{\exp[b(\cos(2\Theta)+1]}{\text{erfi} (\sqrt{2b})},
\label{Von mises}
\end{equation}
where $b$ is the concentration parameter associated with the Von Mises distribution and $\Theta $ is the mean orientation of fibres (for a graph of the variation of the dispersion parameter $\kappa $ with the concentration parameter $b$, see \citet{Gasser06}).

\begin{figure}[!ht]
\centering
\subfigure[$\kappa=0.0085$]{\includegraphics[width=0.4\textwidth,angle=90]{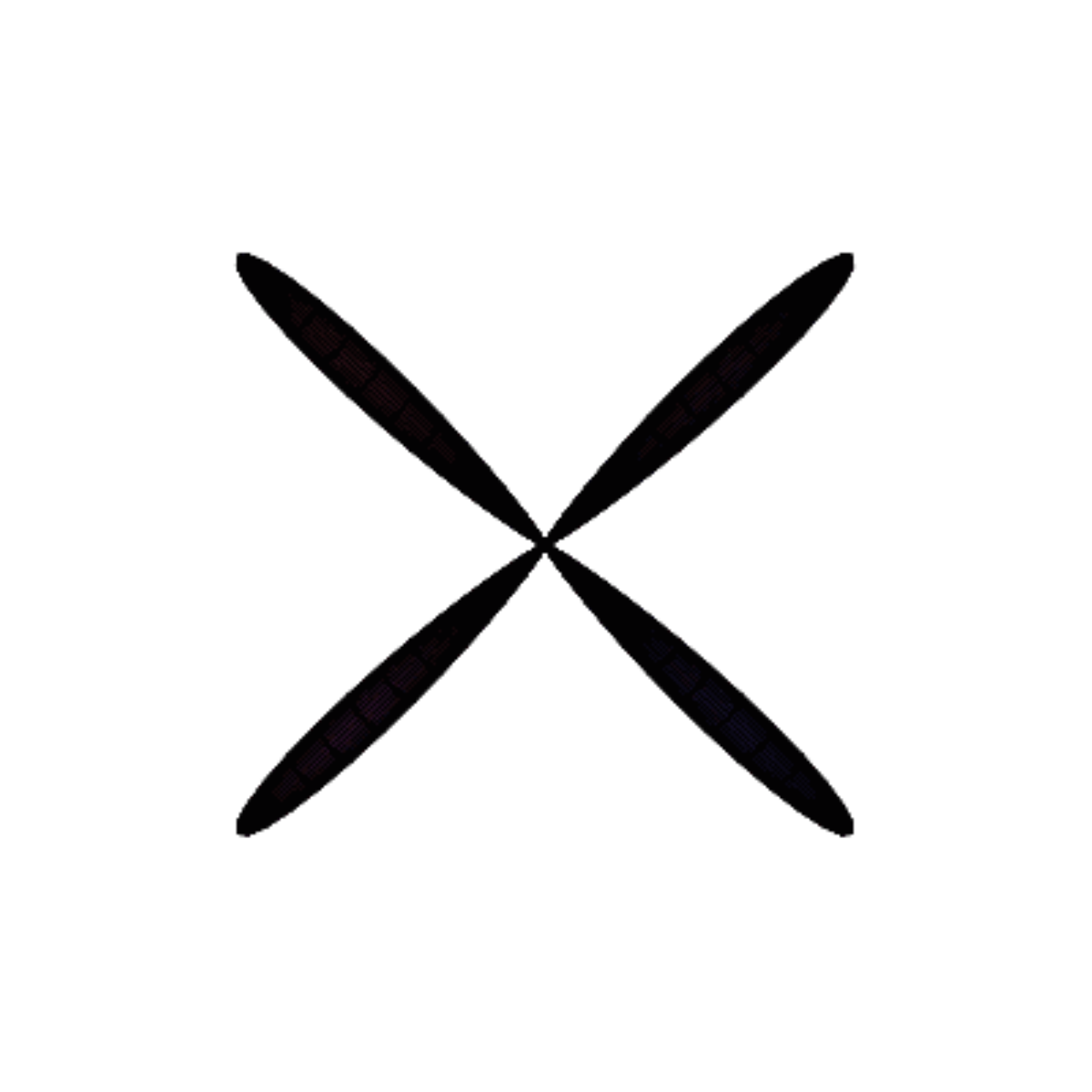}}
\subfigure[$\kappa=0.25$]{\includegraphics[width=0.4\textwidth,angle=90]{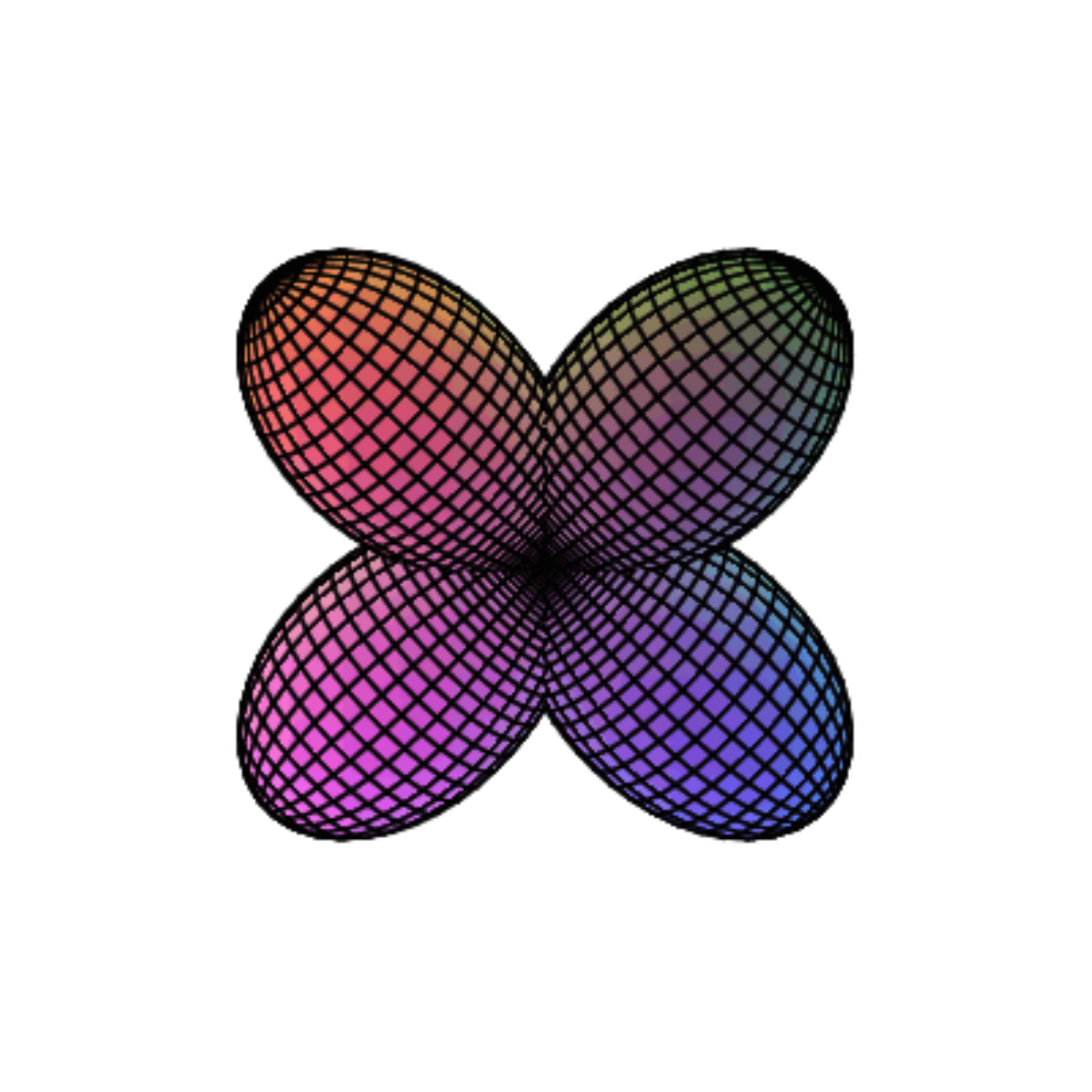} }
\subfigure[$\kappa=0.33$]{\includegraphics[width=0.4\textwidth, angle=90]{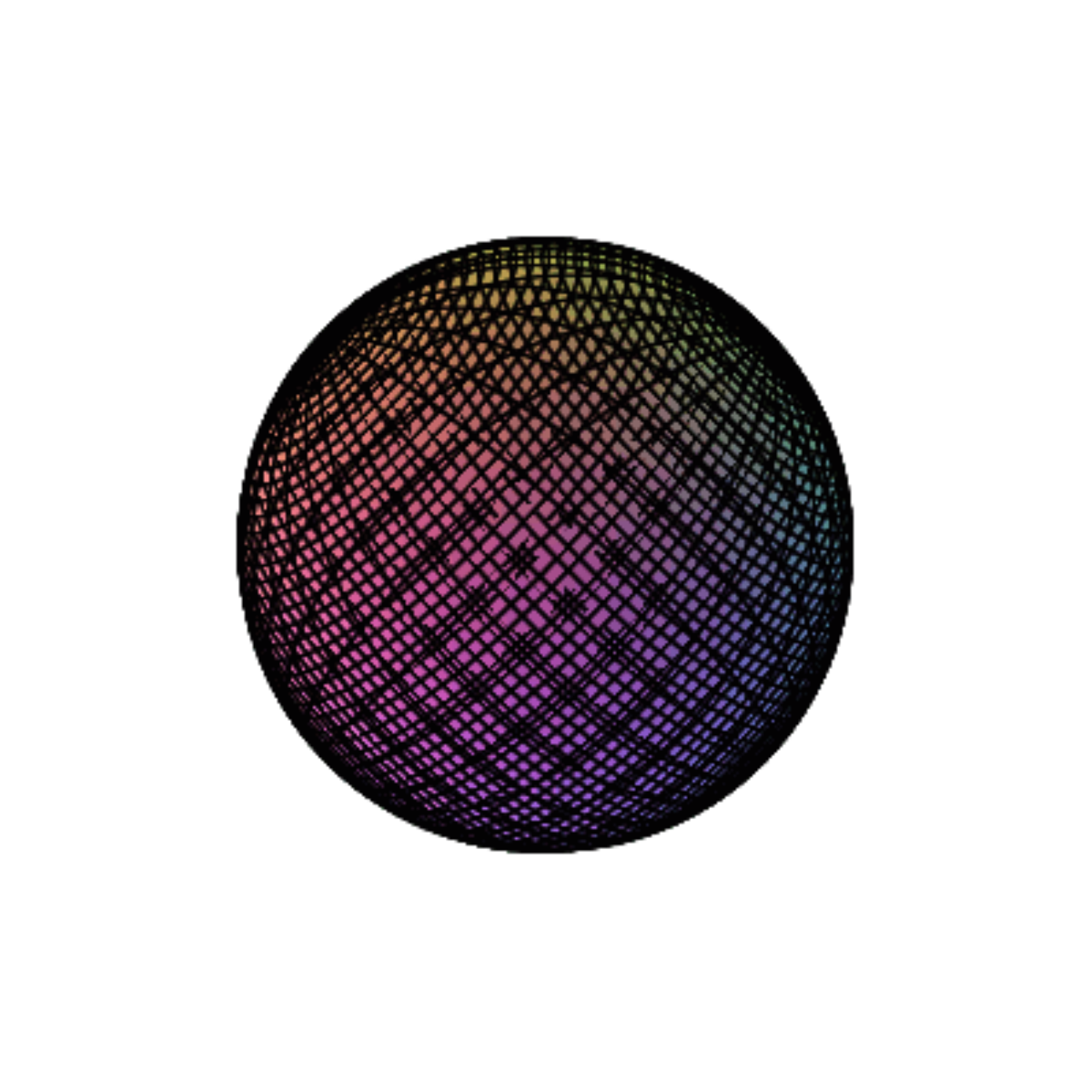} }
\caption{Three dimensional representation of the orientation of collagen fibres.\label{fig:3D_kappa}}
\end{figure}

The parameters $b$ and $\theta$ were evaluated using the \emph{mle} function in MATLAB. Analogous to least squares curve-fitting, the maximum likelihood estimates (MLE), is the preferred technique for parameter estimation in statistics.
Then $\kappa $ is calculated by numerical integration of the integral given by \citet{Gasser06},
\begin{equation}
\kappa=\frac{1}{4}\int^\pi_0{\rho(\Theta)\sin^3\Theta d\Theta}.
\label{kappa}
\end{equation}
The structural parameter $\kappa$ describes the material's degree of anisotropy. 
It must be in the range $0\le \kappa \le 1/3$: 
the lower limit, $\kappa  = 0$, relates to the ideal alignment of collagen fibres and 
the upper limit, $\kappa=1/3$, relates to the isotropic distribution of collagen fibres. 
Fig.~\ref{fig:3D_kappa} is a 3D graphical representation of the orientation of collagen fibres for different values of $\kappa$.

The fibres were assumed to form an interweaving lattice structure as first postulated by \citet{Ridge66} and shown in Fig.~\ref{weave}. 
These authors suggested that the mean angle of the two families of fibres indicates the direction of the Langer lines. More recent in vitro \citep{Jor11} and in vivo \citep{Ruvolo07} studies have also supported this hypothesis. The lattice structure proposed by \citet{Ridge66} is an idealised one, and the adoption here of the dispersion factor creates a more realistic scenario. 
Finally, recall that the two families (directions) of fibres are assumed to have a common dispersion factor.


\subsection{Finite element representation}

\begin{figure}[!ht]
\centering
 \includegraphics[width=0.8\textwidth]{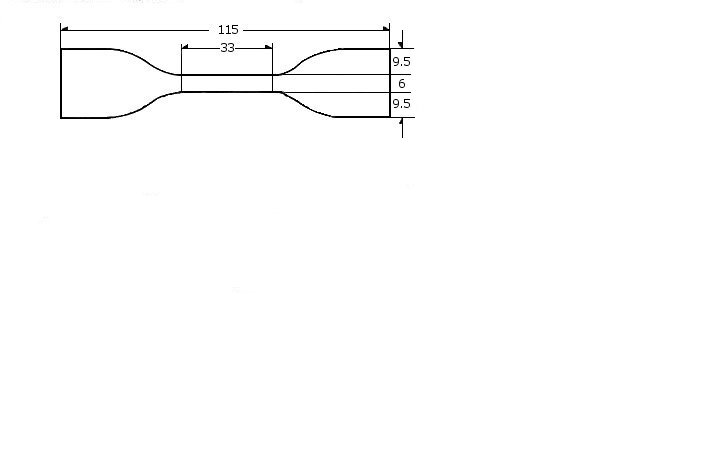}
\caption{Dimensions of test specimen (mm).\label{die}}
\end{figure}

An FE computation was used to simulate uniaxial tensile tests of human skin which were described in a previous publication \citep{NiAnnaidh11a}. The simulation was carried out for three samples: parallel, perpendicular, and at 45$^\circ$ to the Langer lines. The test samples were of the dimensions shown in Fig.~\ref{die}. The length of the unclamped specimen was 68 mm and the thickness was 2.25 mm. 
1512 reduced integration hybrid hexahedral (C3D8RH) elements were used for the mesh. The numerical analyses were performed using the static analysis procedure in ABAQUS/Standard. Displacement was applied through a smooth amplitude boundary condition, and the top and bottom of the sample were encastred to represent the clamping of samples. The material model used was the anisotropic GOH model, which is an internal material model in ABAQUS.


\section{Results}
\subsection{Structural parameters}

\begin{figure}[!ht]
\centering
 \includegraphics[width=1.0\textwidth]{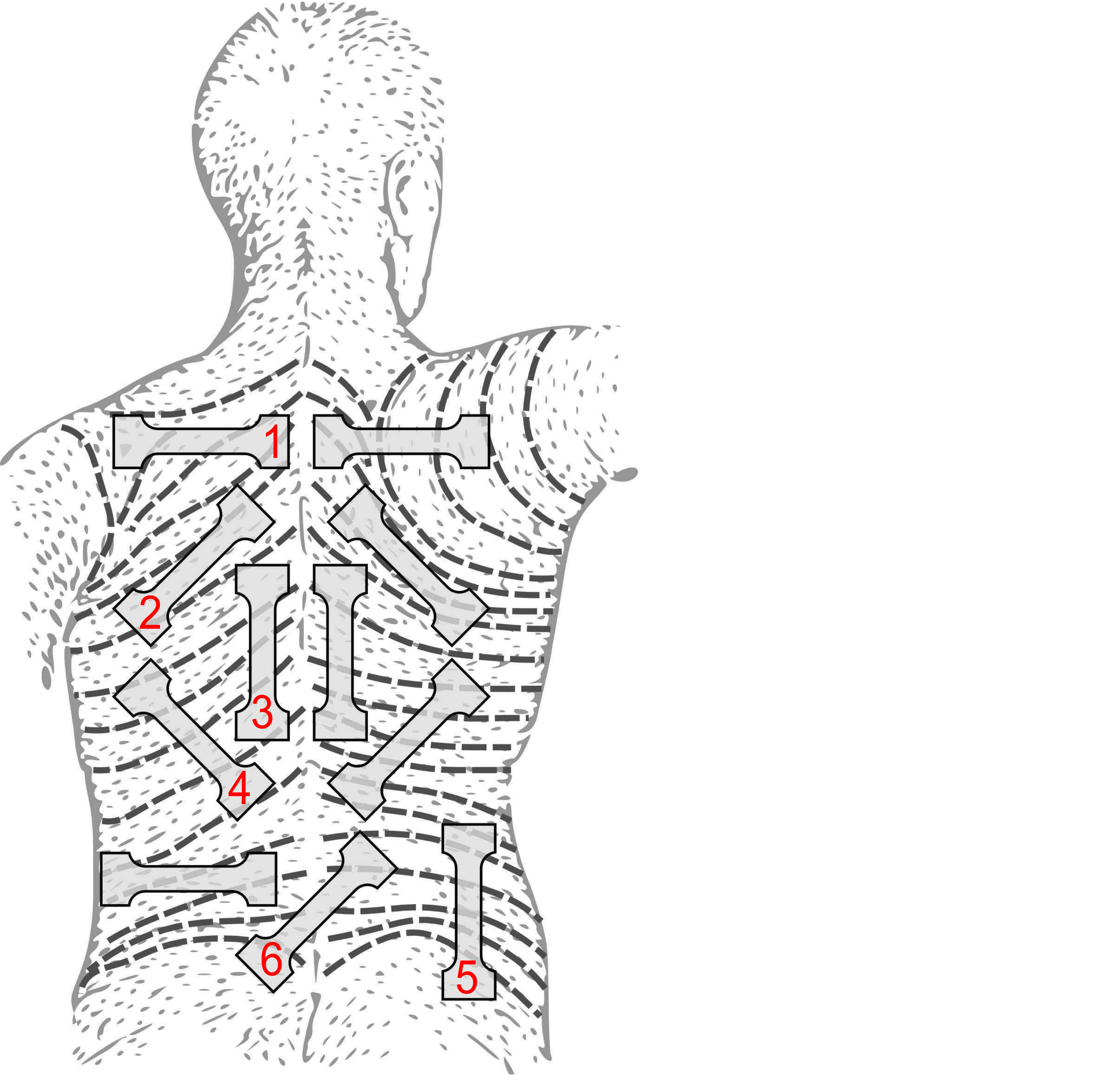}\caption{Location of tensile test samples shown in Table~\ref{orientations} (figure amended from \citep{Langer78}).\label{orientationsfig}}
\end{figure}

As expected, it was found from the histology that the collagen fibres were locally orientated. This meant that each sample had a different mean orientation and fibre dispersion. Table~\ref{orientations} tabulates the results of 12 different human skin samples, with different orientations, which were procured from the backs of two different subjects\footnote{It should be noted here that these results have been obtained using the \emph{mle} method described in Section~\ref{Dispersion}. An alternative, but less reliable method exists whereby data is clustered into two intersecting fibre families using an agglomerative clustering algorithm such as was performed in \citet{NiAnnaidh11a}.}. See Fig.~\ref{orientationsfig} for details of specimen location and orientation. The preferred orientation, $\widehat \Theta$, refers to the bisector of the two families of fibres. The $95\%$ Confidence Interval of the mean of these images ranged from $1.16^\circ-2.77^\circ$, thereby indicating that the preferred orientation does not change significantly over this small area. The results of our validation revealed that the automated process differed from the manual segmentation by an average of 5$^\circ$$\pm$4$^\circ$, giving us further confidence in the validity of our technique.

A Pearson correlation test was carried out to test for a correlation between the measured preferred orientation obtained through histology and the perceived orientation of Langer lines (the natural lines of tension in the skin). The orientation of Langer lines was assessed using generic maps, described further in \citet{NiAnnaidh11a}.The correlation was deemed to be significant (P$<$0.001) with an R$^{2}$ value of 0.95. This shows that the Langer lines have an anatomical basis, a point which had previously been suggested but until now had not been quantitatively assessed.


\subsection{Constitutive parameters}
\label{Parameters}

It was assumed that the orientation of collagen fibres is symmetric about the axis of applied stress. 
In reality, this is not always the case; however some assumptions must be made in order to ensure that the constitutive relations remain practical for numerical implementation. 
In particular, as explained in Section \ref{Gasser}, this assumption leads to a homogeneous deformation of the sample, and in turn, to an explicit stress-strain solution.
In this section, three illustrative examples from Table~\ref{orientations} (highlighted) have been chosen for further investigation. 
Two of these examples have been chosen because they are the samples that are closest to being symmetrical about the axis of applied stress. The third sample has been chosen to illustrate how its behaviour can be modelled using FE analysis and is described further in Section~\ref{FE} 

\begin{figure}[!ht]
\centering
\subfigure[]{\includegraphics[width=0.8\textwidth]{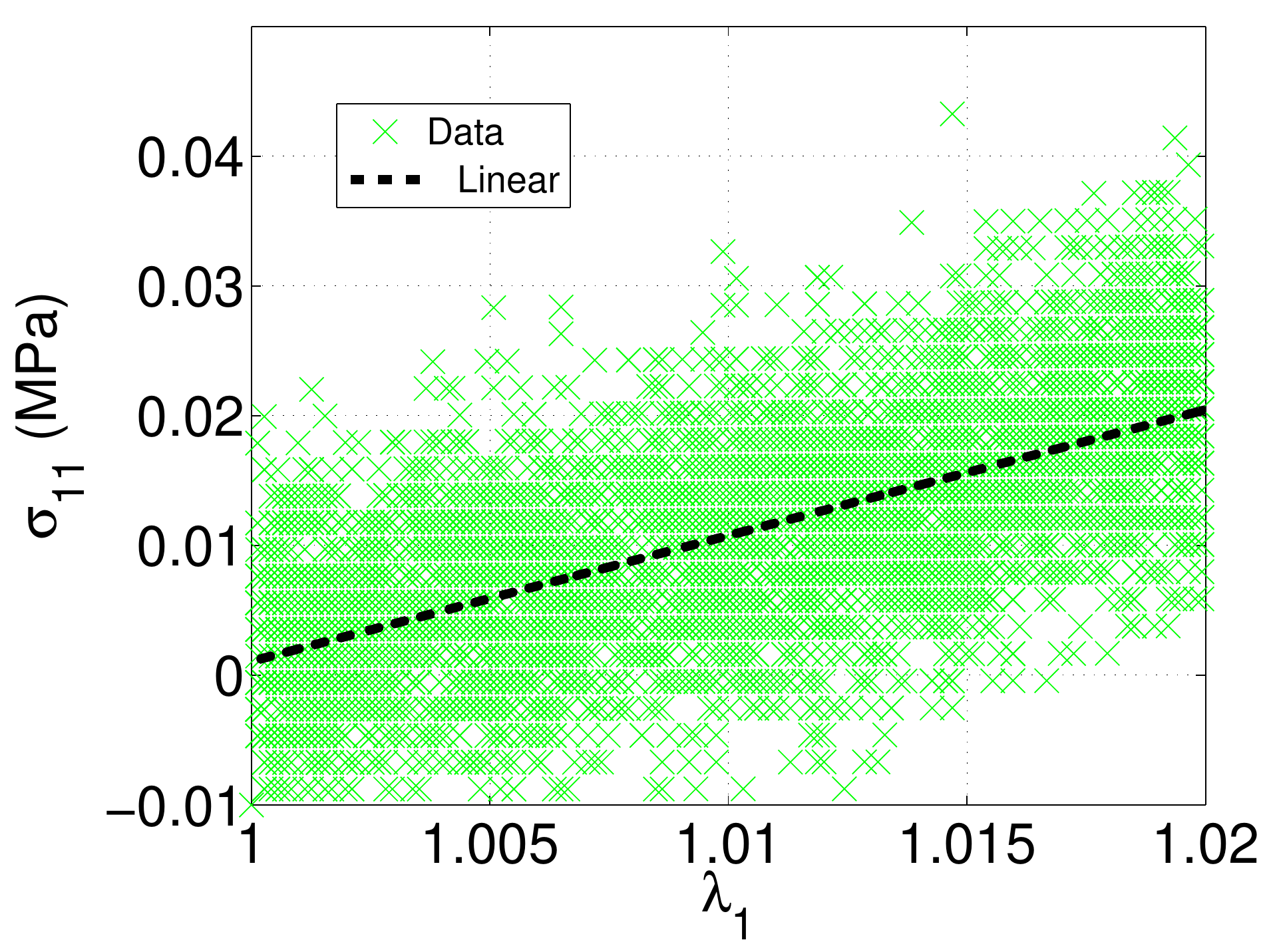} }
\subfigure[]{\includegraphics[width=0.8\textwidth]{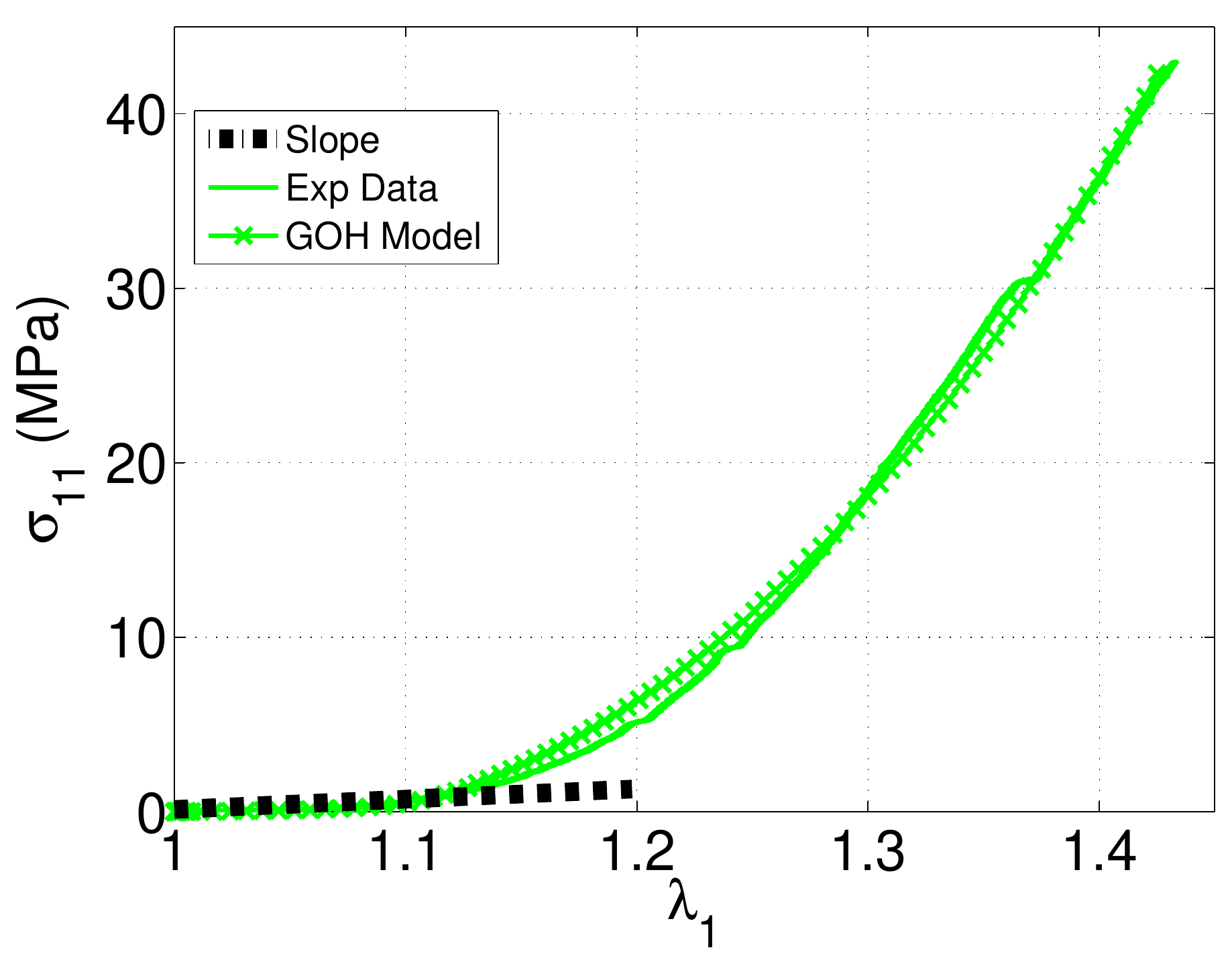} }
\caption{Nonlinear curve fitting to obtain the constitutive parameters: $\mu$ and $k_1$ are related to the early (infinitesimal) stress-strain part of the graph, see (a); $k_2$, to the rest of the curve.}
\label{fig-tensile}
\end{figure}

The constitutive parameters for the GOH model are obtained by using Equations \eqref{sigma1} and \eqref{sigma2}, obtained in Section~\ref{Gasser}. 
When linearized in the neighbourhood of small strains, $\lambda_i \simeq 1 + \epsilon_i$, say, we find that they read as follows,
\begin{align}
\sigma_{11} &= 4\mu[1 + 2k_1(1-3\kappa)^2\cos^4\gamma]\epsilon_1 + 2\mu [1 + 4k_1(1-3\kappa)^2\sin^2\gamma\cos^2\gamma]\epsilon_2, \label{linear1} \\
 0 &= [1+4k_1(1-3\kappa)^2\cos^2\gamma\sin^2\gamma]\epsilon_1 + 2[1 + 2k_1(1-3\kappa)^2\sin^4\gamma]\epsilon_2. \label{linear2}
\end{align}  
These expressions reveal that the constitutive parameters $\mu$ and $k_1$ are related to the early stages of the tensile tests, whilst $k_2$ is a stiffening parameter, related to the latter (nonlinear) stages of the tensile tests. 
By solving \eqref{linear2} for $\epsilon_2$, and substitution into \eqref{linear1}, we find the linear stress-strain relation $\sigma_{11} = E_1 \epsilon_1$, where $E_1$ is the infinitesimal Young modulus in the 1-direction, found here as
\be
E_1 = \dfrac{3+8k_1(1-3\kappa)^2(1-3\cos^2\gamma\sin^2\gamma)}{1+2k_1(1-3\kappa)^2\sin^4\gamma}\mu,
\label{E}
\en
(which is consistent with the formula $E = 3\mu$ in linear isotropic ($\kappa=1/3$) incompressible elasticity).
Hence, by plotting the values of $\sigma_{11}$ for the early part of the tests (first 1000 data say, corresponding to a tensile stretch of less than 2\%), we can determine $E_1$ by linear regression analysis, see Fig.\ref{fig-tensile}(a).
Here we have plotted the `parallel' sample highlighted in Table~\ref{orientations} data for which was collected from tensile tests of human skin samples \citep{NiAnnaidh11a}.

Once $E_1$ is determined,  $\mu$ can be expressed in terms of $E_1$ and $k_1$ using Eq.~\eqref{E}.
Then the remaining material parameters $k_1$ and $k_{2}$ are found through the nonlinear least squares fitting with experimental test data of Equation \eqref{sigma1}, subject to the definition of $\lambda_2$ in terms of $\lambda_1$ given by Equation \eqref{sigma2}.
The data fitting was performed using the \emph{lsqnonlin} MATLAB routine in the Optimisation Toolbox where the objective function, $Err(k)$, was given as

\begin{equation}
Err(k)=\displaystyle\sum\limits_{i=1}^n (y_i^{exp} - y_i^{model(k)})^2
\label{objective}
\end{equation}

Where $n$ is the number of experimental data points, $y_i^{exp}$ is the experimental value and $y_i^{model(k)}$ is the value predicted by the model using the current material parameters, $k$.

Non-linear optimisation procedures are often sensitive to the initial starting point provided by the user \citep{Ogden04}. In our case, the initial estimate for $k_1$ was found by calculating the slope of the non-linear part of the stress-stretch curve. Since we have shown that $k_1$ is related to the stiffening stage of the tensile test, our initial estimate therefore has a physical meaning. Furthermore, the initial estimates of both $k_1$ and $k_2$ were varied over a large range (0-1e$^6$ for $k_1$ and 0-1e$^3$ for $k_2$) and lead to the same set of optimal parameters each time, illustrating that the results of the optimisation procedure are not sensitive to this initial estimate. 

\begin{figure}[!ht]
\centering
\includegraphics[width=0.8\textwidth]{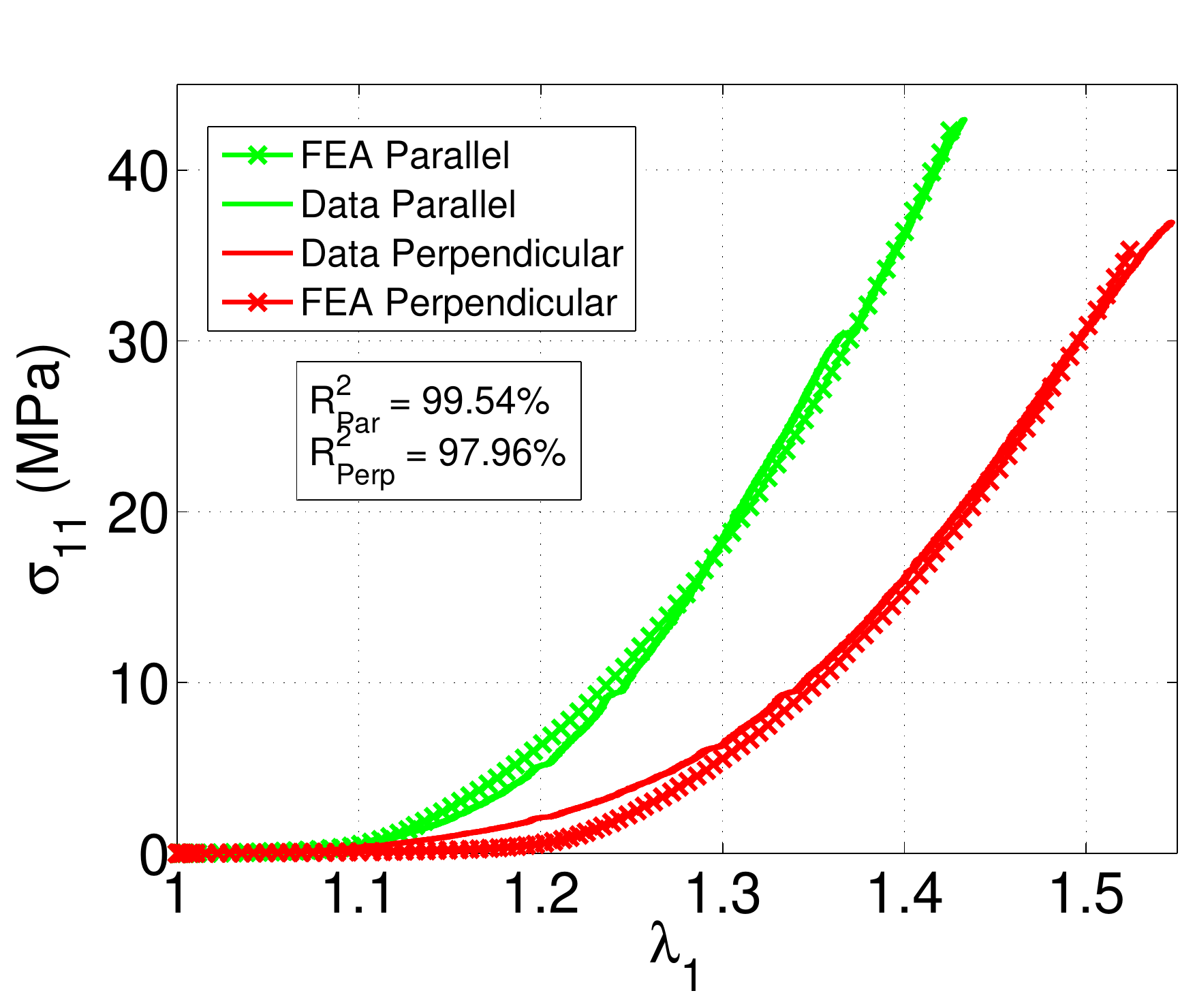} 
\caption{Comparison of sample parallel to Langer lines and perpendicular to Langer lines.}
\label{ParPerp}
\end{figure}

The results of our optimisation procedure gave us a value of 243.6 for $k_1$ and 0.1327 for $k_2$, with an $R^2$ of $99.5\%$. Fig.\ref{fig-tensile}(b) shows the GOH model fit to the experimental data. It can be seen that these material parameters provide an excellent fitting to the `parallel' sample, at least from a \emph{descriptive} point of view. Turning now to the \emph{predictive} capabilities of the GOH model, we examine a sample `perpendicular' to the Langer lines. In Fig.~\ref{ParPerp} we use the material parameters $k_1$ and $k_2$ obtained through least squares fitting, $\mu$ calculated by linear regression and Eq.~\eqref{E}, coupled with the unique structural data for the `perpendicular' sample in Table~\ref{parameters}. We compare the model prediction for a tensile test occurring perpendicular to the Langer lines to the experimental data. The fit remains good with  R$^2$ = 97.96\%. This shows that the model is capable of predicting the behaviour of skin once the structural parameters have been evaluated.


\subsection{Finite element simulations}
\label{FE}

The conventions used by ABAQUS are related to ours through  $C_{10}=\mu/2$, $k'_1=k_1 \mu$ and $k'_2 = k_2$, so that the ABAQUS parameters used were  $C_{10} = 0.1007$ MPa, $k'_1 = 24.53$ MPa and $k'_2 = 0.1327$. 

\begin{figure}[!ht]
\centering
\subfigure[]{\includegraphics[width=0.5\textwidth]{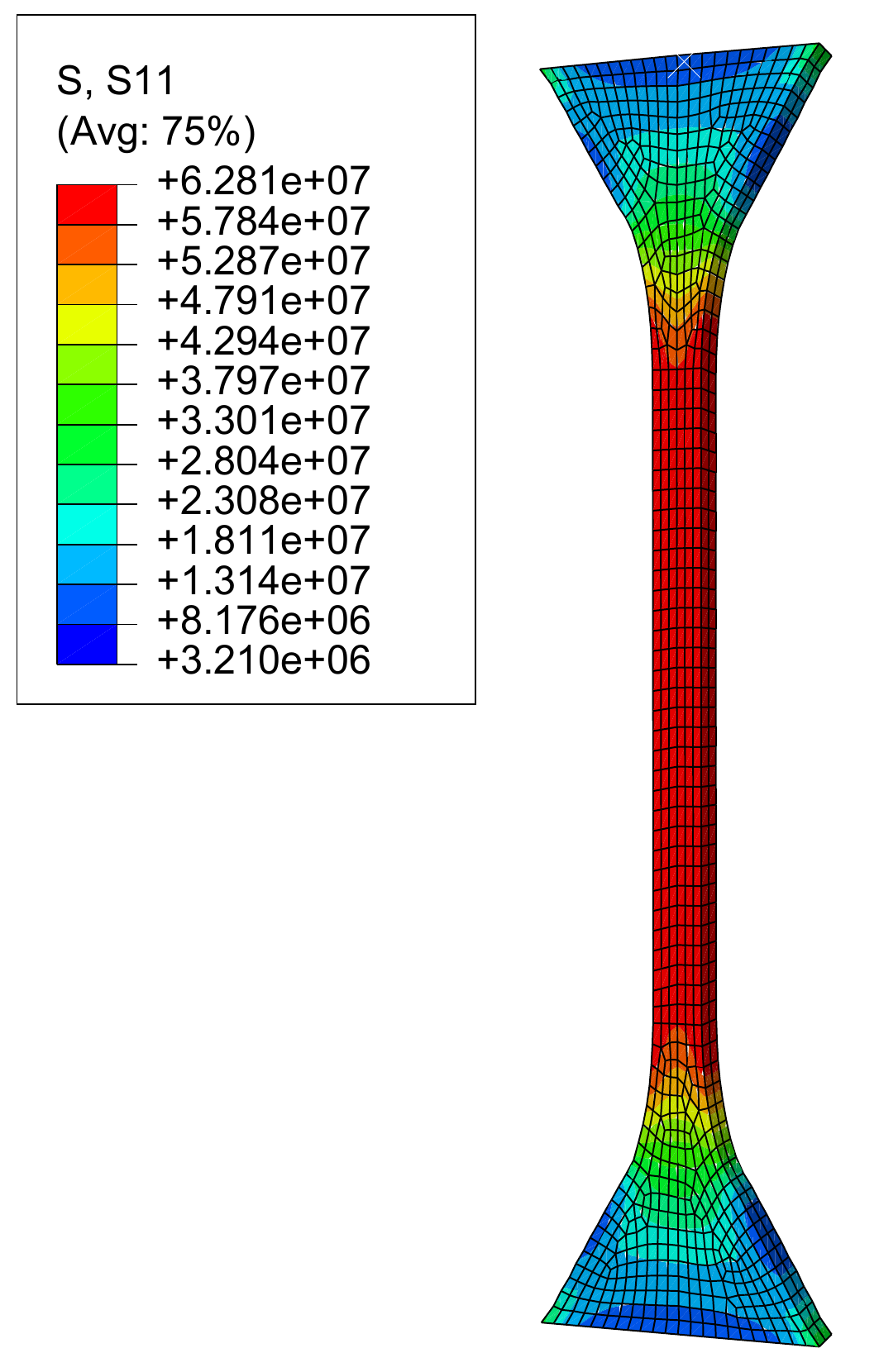} }
\subfigure[]{\includegraphics[width=0.365\textwidth]{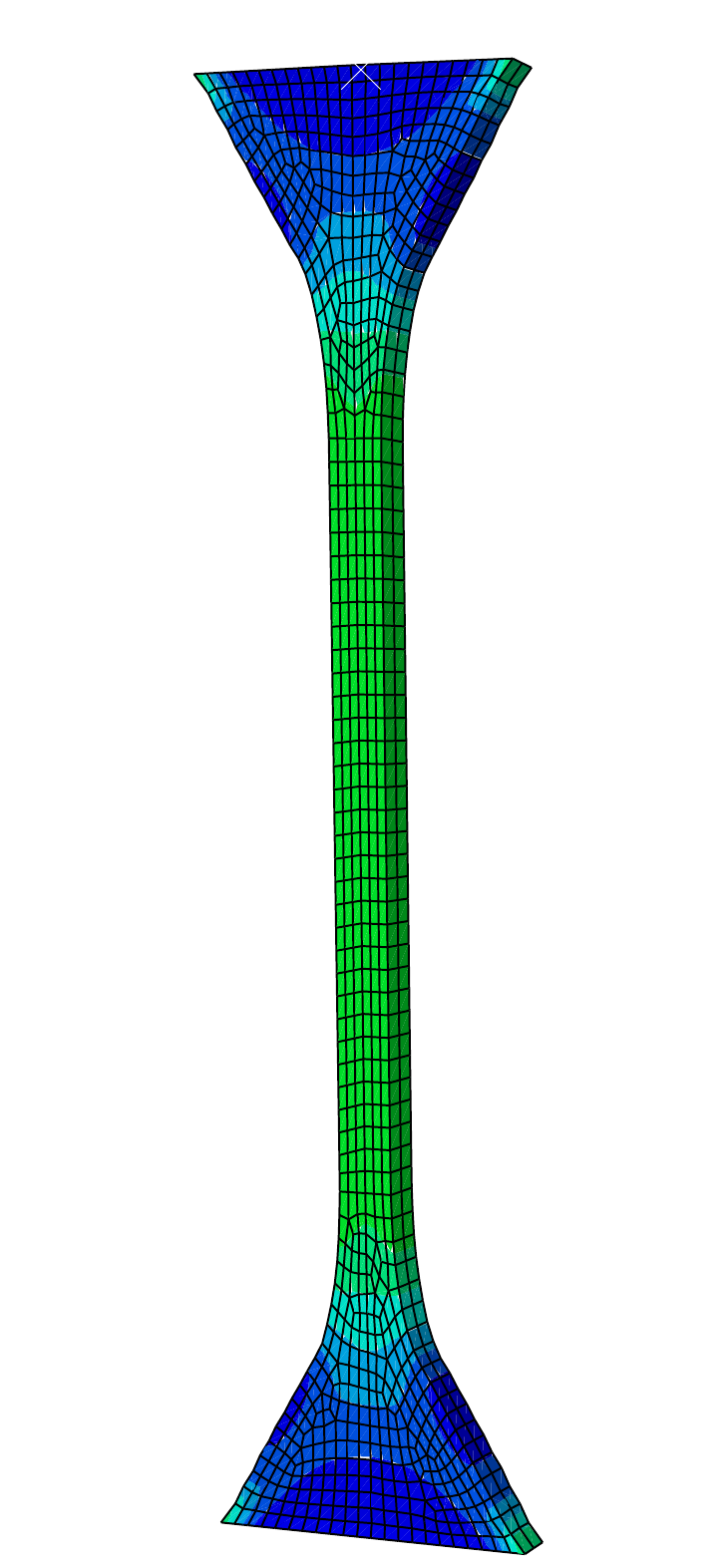} }
\caption{Cauchy stress in Pa of sample strained by 50\% (a) Parallel to the Langer lines (b) Perpendicular to the Langer lines.\label{deformed}}
\end{figure}

The FE simulation results of the uniaxial tensile tests for both the parallel and perpendicular samples were identical to the analytical solution. The results were independent of both mesh density and element type. Fig.~\ref{deformed} shows the Cauchy stress distribution across the parallel and perpendicular samples at the end of the test. The large difference in magnitudes between the two is due to the variation in the mean orientation of fibres. Note the uniform distribution of stress in the middle section of the test samples, thanks to the dog-bone shape of the specimen, and the symmetry of fibres about the axis of applied stress. 

\begin{figure}[!ht]
\centering
\subfigure[]{\includegraphics[width=1\textwidth]{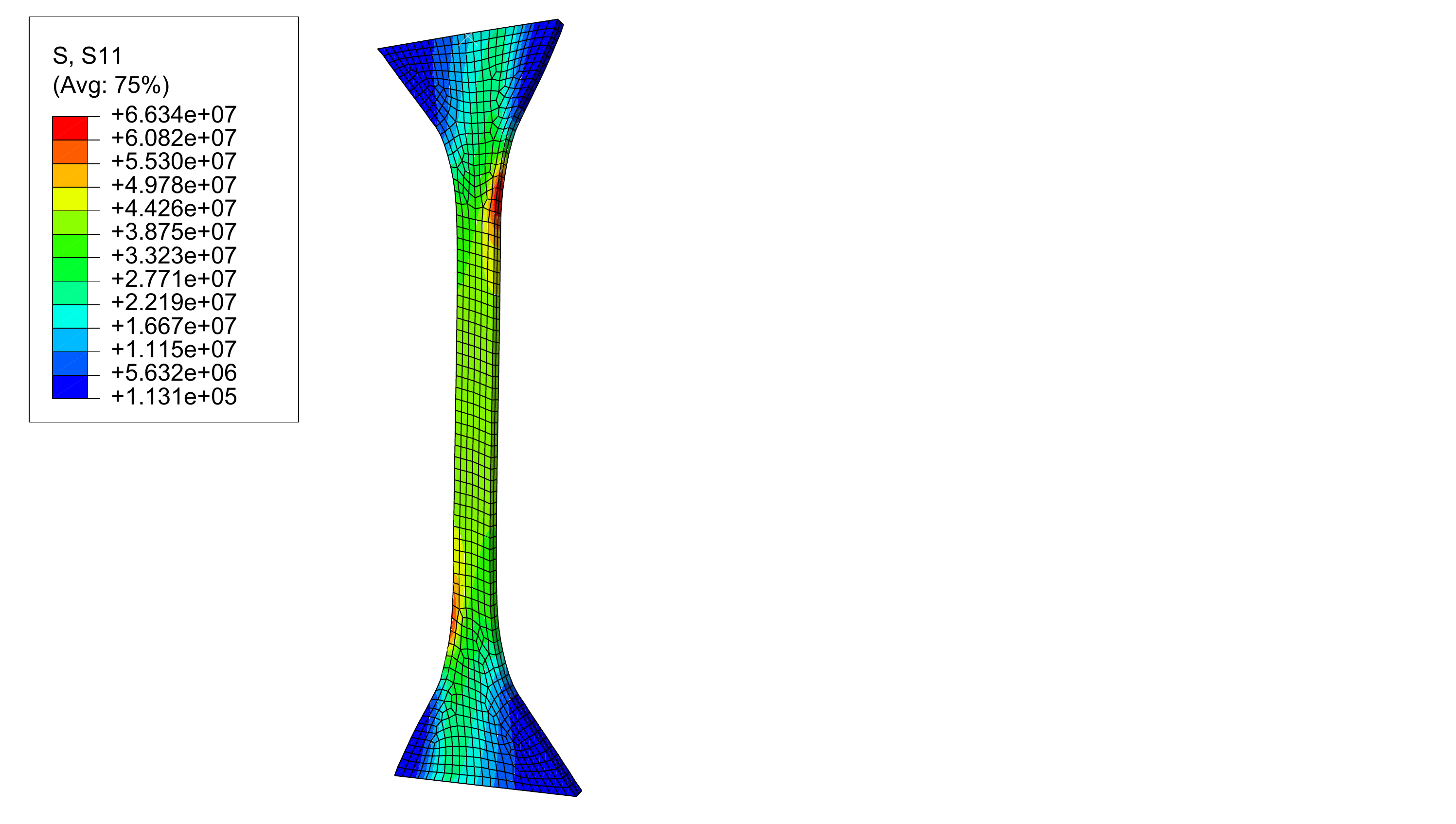} }
\subfigure[]{\includegraphics[width=1\textwidth]{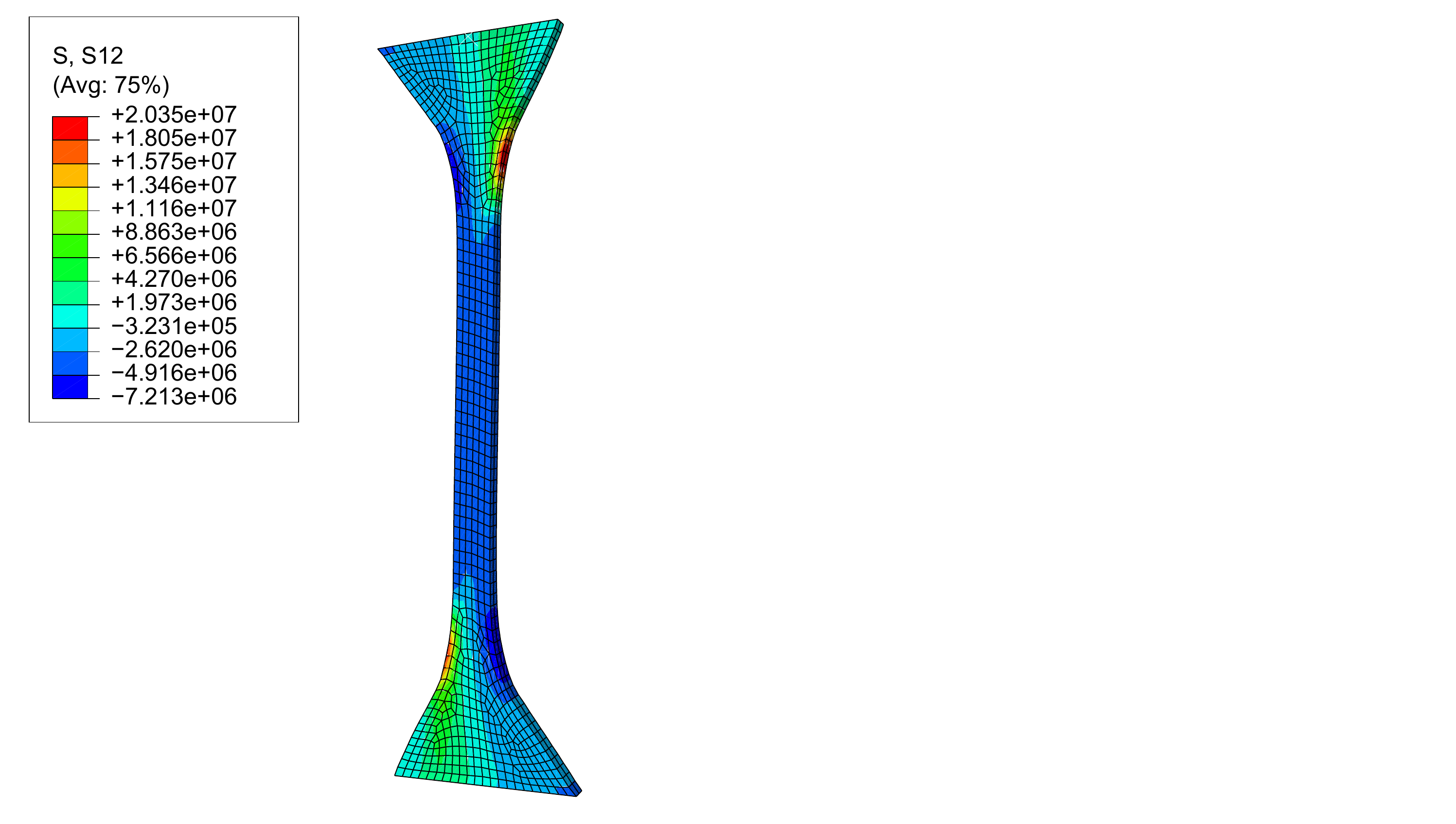} }
\caption{Cauchy stress in Pa of a non-symmetric sample strained by 30\% (a) $\sigma_{11}$ (b) $\sigma_{12}$}
\label{shear}
\end{figure}

As discussed in Section~\ref{Parameters}, for ease of determining the material parameters, it was assumed that the orientation of collagen fibres is symmetric about the axis of applied stress. This is for an idealised scenario only, where one knows the exact orientation of collagen fibres prior to testing, and can therefore apply the stress in this orientation. However, with the model parameters that have now been determined, a FE simulation can provide results for samples where the collagen fibres are not symmetric about the axis of applied stress. Examining the non-symmetric example in Fig.~\ref{shear}(a), we can see a non-uniform distribution of stress throughout the test specimen. A local magnification of stress occurs near the neck regions of the test specimen. This is due to the non-symmetry of collagen fibres about the axis of applied stress and makes this problem a much more complicated one to solve analytically. The presence of significant levels of shear in Fig.~\ref{shear}(b) (which is absent from both the parallel and perpendicular samples) indicates further the effect of this non-symmetry on the sample response.

\begin{figure}[!ht]
\centering
\includegraphics[width=1\textwidth]{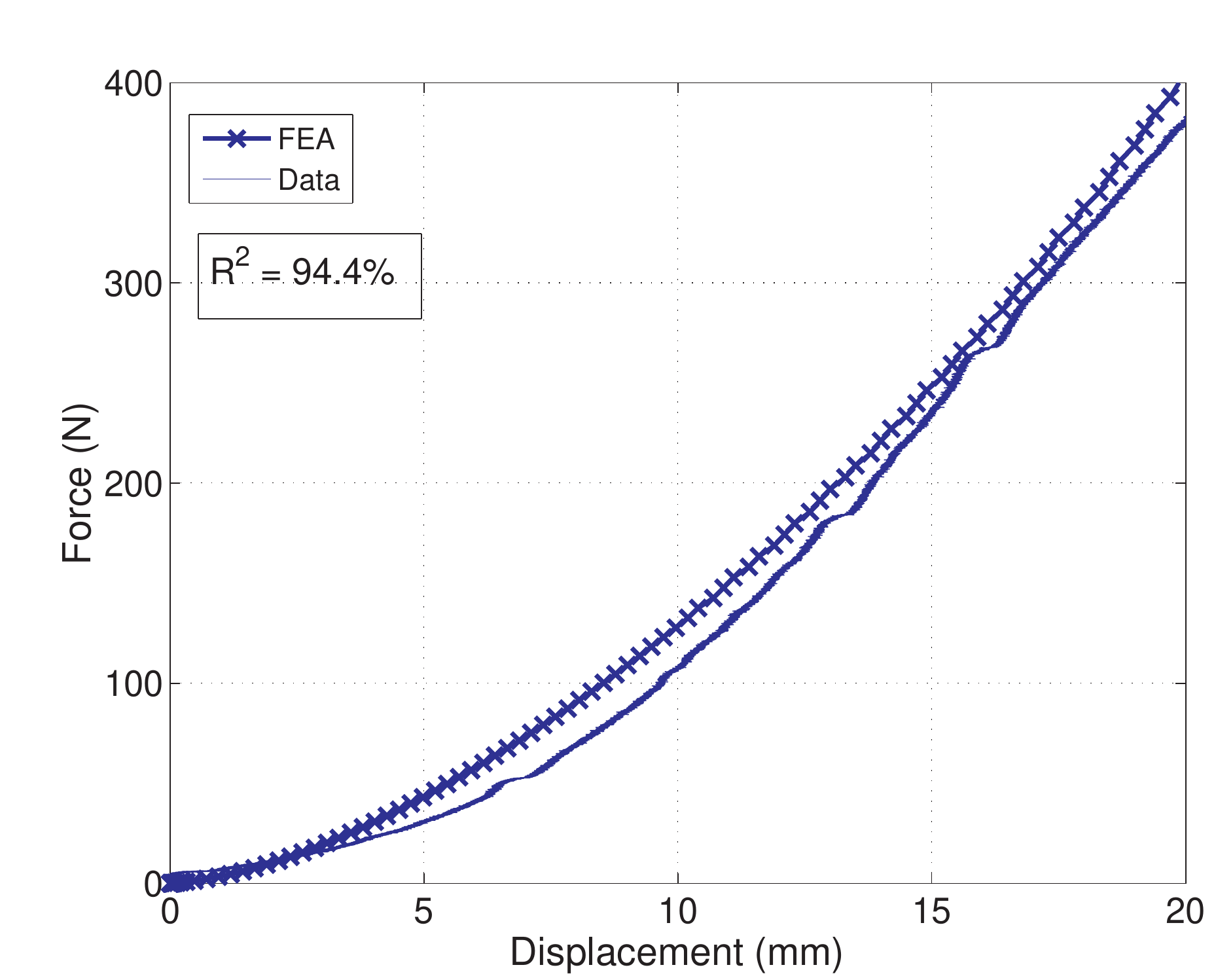} 
\caption{Comparison between predicted model response and experimental force-displacement data for a non-symmetric sample.}\label{shearforce}
\end{figure}

Because the stress distribution in this sample is non-uniform, to examine the \emph{predictive} capabilities of the GOH model here, we must plot the experimental force-displacement data against the values predicted by ABAQUS for a node at the top of the test sample, see Fig.~\ref{shearforce}. Again, we have found that the model predicts the behaviour well with an $R^2$ of $94.4\%$, showing that the GOH model is capable of predicting the anisotropic response of human skin.


\section{Discussion}

For this paper, histology slides in three different planes were examined; however, after capturing the images from all three planes it was observed that three times as many cross-sectioned fibres were present in the plane normal to the epidermis, therefore an assumption was made to ignore the fibres normal to the epidermis. Hence, the further analysis of the samples was restricted to the plane parallel to the epidermis alone, making this a 2D analysis. Physically however, there are a percentage of fibres that run normal to the epidermis and this information has not been captured here. Furthermore, unloaded collagen fibres have a crimped nature: Their relative orientation may vary with respect to the epidermal plane. Here we have tried to overcome this limitation by taking an average measure over six levels of depth spanning 30$\mu$m,with the view that the general orientation of the collagen fibres are still captured. Ideally, a full 3D analysis of the dermal structure could consider the effects of collagen crimping. A 3D analysis was beyond the scope of this paper, but the current technique could be extended by creating a montage of overlapping images through the thickness of the dermis, as described in \citet{Jor11}, therefore turning the 2D analysis into a 3D analysis.

A further limitation of this technique is that we have assumed that the structure of the skin biopsy removed is representative of the entire tensile test sample. In reality, the structure, and therefore the properties of skin may vary considerably over a small area. The variation of the mean fibre orientation over the volume of the biopsy was quantified, however, the size of the biopsy is very small relative to the tensile test sample and we cannot infer that the variation would be negligible. At a minimum, future studies should excise a number of biopsies along the length of the test specimen to investigate the variation of the structure.  

While this technique has been described as `automated', there are still a number of `manual' steps that must first be performed. The first manual task is the histological staining, combined with the mounting of skin biopsies. The collagen detection process demands a high quality of histological staining for the method to be successful and therefore, care must be taken to follow standard procedures carefully. The second `manual' task is the image acquisition phase. Modern `slide scanners' automatically scan multiple slides at once meaning that tedious image acquisition techniques using a manual microscope are no longer necessary, however images must still be captured from the `digital slide'.

While this technique has provided quantitative structural data of human skin, it can, of course, only be applied \emph{in-vitro}. Considering the effect that the mean orientation of collagen fibres has on the mechanical response of skin, the development of \emph{in-vivo} methods for establishing the orientation of fibres is of the utmost importance. Advanced imaging techniques such as ultrasonic surface wave propagation may eventually provide real-time, \emph{in-vivo} structural data.

It should be noted that the nonlinear curve fitting technique rests only on measurements of $\lambda_1$ and $\sigma_{11}$, and that $\lambda_2$, along with $k_1$ and $k_2$ are obtained during the simultaneous optimisation of  Eq.~\eqref{sigma1} and Eq.~\eqref{sigma2}. Ideally, a more complete analysis would include, compare, and contrast experimental data for $\lambda_2$ and/or $\lambda_3$. An extensive experimental data set would include planar biaxial tests with in-plane shear and separate through thickness shear tests \citep{Flynn11}\citep{Holzapfel09}\citep{Jor11b}, however in the absence of these advanced testing protocols tensile tests coupled with a histological study of the collagen fibre alignment can be used for reasonable determination of material parameters \citep{Holzapfel06}. Of course non-uniqueness of ‘optimal’ material parameters is an intrinsic problem in non-linear fitting. It is possible that the optimisation procedure finds a local minimum and assumes that this is the global minimum \citep{Ogden04}. To ensure that the optimisation procedure is providing a unique set of material parameters a number of checks are available: The properties of the Hessian matrix can be investigated at the optimum \citep{Gamage11}\citep{Lanir96}, or alternatively, one can plot the objective function as a function of the varying material parameters. In this case the objective function was investigated and the plot (not reproduced) shows that our procedure calculates a global minimum and not merely a local minimum.

In this study we have developed a simple automated process which can detect the orientation of collagen fibres. This technique can be easily implemented in MATLAB and can be adapted to detect other biological features, such as certain cells, leading to applications in diagnostics. We have applied this technique to skin biopsies and provided new quantitative data on the orientation of collagen fibres in the human dermis. 
So far, the availability of accurate structural data has lagged behind the progress of anisotropic constitutive modelling. Here we have provided the structural data required to accurately make use of advances in constitutive modelling, and help fill the void of experimental data. The model parameters of the GOH model have been evaluated for skin using experimental data from the same skin samples. These sets of parameters will provide invaluable data for those wishing to model the anisotropic behaviour of skin. Finally, an FE simulation of a uniaxial tensile test on three separate  human skin samples was performed which predicted the the response of these three samples well. We have illustrated that the Gasser-Ogden-Holzapfel model can successfully model the anisotropic behaviour of human skin and that it can be implemented in ABAQUS with ease. 


\section{Acknowledgements}

The authors acknowledge gratefully the advice and assistance of Mr. Ciaran Driver, Dr. Michael Curtis and Prof. Marie Cassidy, of the Office of the State Pathologist (Ireland), in the area of histology. 
This research was supported by a Marie Curie Intra European Fellowship within the 7$^{th}$ European Community Framework Programme, awarded to MD; by the Irish Research Council for Science, Engineering and Technology; by the Office of the State Pathologist (Irish Department of Justice and Equality); and by the Ile-de-France region. G.S. is supported by the PRIN 2009 project ``Matematica e
meccanica dei sistemi biologici e dei tessuti molli".




\section{Tables \& Figures}

\newcolumntype{x}[1]{>{\centering\hspace{0pt}}p{#1}}
\begin {table}[!ht]
\centering
\caption{Local mean orientation of fibres and dispersion factor. Orientations are given with respect to the axis perpendicular to the axis of applied stretch. The three samples highlighted are those taken as illustrative examples for further analysis. (Note that the data given is axial data i.e. it represents undirected lines and does not distinguish between $\theta $ and $\pi +\theta$ \citep{Jones06} e.g. the orientation of 0\r{ } and 180\r{ } are equivalent).\label{orientations}}
\begin{tabular}{|x{1cm}|x{1.5cm}|x{1.5cm}|x{2.5cm}|x{2.5cm}|x{3.5cm}|}
\hline
Age			&Gender			&Location	 &Orientation of Langer lines	& Preferred Orientation $\widehat \Theta$	& Dispersion factor $\kappa $ \tabularnewline
			&				&		 &$^\circ$			&$^\circ$					&\tabularnewline				
\hline
\multirow{6}{*}{81} &\multirow{6}{*}{Female}	&5		&0/180				& 174$\pm $3				& 0.1306$\pm$0.0054  \tabularnewline
			&				&4		&0/180				& 20$\pm $3				& 0.1439$\pm$0.0088  \tabularnewline
			&				&6		&45				& 38$\pm $7				& 0.1314$\pm$0.0054  \tabularnewline
			&				&1		&45				& 46$\pm $8				& 0.1675$\pm$0.0023  \tabularnewline
			&				&\multicolumn{1}{>{\columncolor[gray]{0.8}}c}{2}		&\multicolumn{1}{>{\columncolor[gray]{0.8}}c}{90}	& \multicolumn{1}{>{\columncolor[gray]{0.8}}c}{88$\pm $8}				&\multicolumn{1}{>{\columncolor[gray]{0.8}}c} {0.1535$\pm$0.0059}		 \tabularnewline
			&				&3		&135				& 121$\pm $5				& 0.1485$\pm$0.0026 \tabularnewline
\hline
\multirow{6}{*}{89} &\multirow{6}{*}{Male}	&5		&0/180			& 178$\pm $4				& 0.1462$\pm$0.0053  \tabularnewline
			&				&\multicolumn{1}{>{\columncolor[gray]{0.8}}c}{4}		&\multicolumn{1}{>{\columncolor[gray]{0.8}}c}{0/180}		& \multicolumn{1}{>{\columncolor[gray]{0.8}}c}{0$\pm $5}	& \multicolumn{1}{>{\columncolor[gray]{0.8}}c}{0.1456$\pm$0.0055}  \tabularnewline
			&				&6		&45			& 61$\pm $4				&0.1289 $\pm$0.0046  \tabularnewline
			&				&1		&45			& 13$\pm $3				& 0.1276$\pm$0.0054  \tabularnewline
			&				&2		&90			& 89$\pm $5				& 0.1009$\pm$0.0085 \tabularnewline
			&				&\multicolumn{1}{>{\columncolor[gray]{0.8}}c}{3}		&\multicolumn{1}{>{\columncolor[gray]{0.8}}c}{135}			&\multicolumn{1}{>{\columncolor[gray]{0.8}}c}{118$\pm $7}	&\multicolumn{1}{>{\columncolor[gray]{0.8}}c} {0.1602$\pm$0.0095} \tabularnewline
\hline
\end{tabular}
\end{table}
\pagebreak

\newcolumntype{x}[1]{>{\centering\hspace{0pt}}p{#1}}
\begin {table}[!ht]
\centering
\caption{ Values obtained through curve-fitting for the parameters $\mu$, $k_{1}$ and $k_{2}$. Also displayed is R$^{2}$, a measure of goodness of fit, and $\kappa$ and $\gamma$ obtained directly through histology. \label{parameters}}
\begin{tabular}{|x{3cm}|x{1.2cm}|x{1.2cm}|x{1.2cm}|x{1cm}|x{2.5cm}|x{1.2cm}|x{1.2cm}|}
\hline
Sample		& $\mu$	 & $k_{1}$	&$k_{2}$	&$\gamma$	&$\widehat \Theta$	&$\kappa$	  &$R^2$\tabularnewline
		&MPa		&               	&              	 &                     &$^\circ$         				&		&$\%$ \tabularnewline
\hline
Parallel		&0.2014	&243.6		&0.1327	&41		&88						&0.1535		&99.54	\tabularnewline
\hline
Perpendicular	&0.2014	&243.6		&0.1327	&41		&0						&0.1456		&97.96	\tabularnewline
\hline
Non-symmetric	&0.2014	&243.6		&0.1327	&41		&118						&0.1602		&94.40	\tabularnewline
\hline
\end{tabular}
\end{table}
\pagebreak

\pagebreak

\pagebreak

\pagebreak

\pagebreak

\pagebreak

\pagebreak

\pagebreak

\pagebreak

\pagebreak

\pagebreak

\pagebreak

\pagebreak

\end{document}